\documentclass[twocolumn]{aastex701}

\usepackage{lineno}
\usepackage{placeins}
\usepackage{wrapfig}
\usepackage{xcolor}
\usepackage{lipsum}
\usepackage[english]{babel}

\usepackage{hyperref}

\begin{document}

\title{ A Not-So-Compact Companion: Massive, Oversize White Dwarf in a Post-Common Envelope Eclipsing Binary}

\author[0009-0001-9841-0846]{Erin Motherway}
\affiliation{Department of Astronomy, University of Wisconsin-Madison, 475 N. Charter St., Madison, WI 53706, USA}
\email{motherway@wisc.edu}

\author[0009-0006-7474-7889]{Evan Linck}
\affiliation{Department of Astronomy, University of Wisconsin-Madison, 475 N. Charter St., Madison, WI 53706, USA}
\email{elinck@wisc.edu}

\author[0000-0002-7130-2757]{Robert D. Mathieu}
\affiliation{Department of Astronomy, University of Wisconsin-Madison, 475 N. Charter St., Madison, WI 53706, USA}
\email{mathieu@astro.wisc.edu}

\author[0000-0001-6977-9495]{Don Dixon}
\affiliation{Department of Physics and Astronomy, Vanderbilt University, 6301 Stevenson Center Lane, Nashville, TN 37235, USA}
\email{don.m.dixon@vanderbilt.edu}

\author[0000-0002-3481-9052]{Keivan G.\ Stassun}
\affiliation{Department of Physics and Astronomy, Vanderbilt University, Nashville, TN 37235, USA}
\email{keivan.stassun@Vanderbilt.Edu}

\author[0000-0001-5228-6598]{Katelyn Breivik}
\affiliation{McWilliams Center for Cosmology and Astrophysics, Department of Physics, Carnegie Mellon University, Pittsburgh, PA 15213, USA}
\email{kbreivik@andrew.cmu.edu}

\author[0000-0003-2025-3147]{Steve Majewski}
\affiliation{Department of Astronomy, University of Virginia, P.O. Box 400325, Charlottesville, VA 22904-4325}
\email{srm4n@virginia.edu}

\author{Onno R. Pols}
\affiliation{Department of Astrophysics/IMAPP, Radboud University, P.O. Box 9010, 6500 GL Nijmegen, The Netherlands}
\email{o.pols@astro.ru.nl}

\begin{abstract}

We provide a detailed characterization of 2M07515777+1807352, a post-common envelope eclipsing binary system with a 10.3 d, nearly, but not quite, circular orbit (e = 0.02). This system consists of a massive white dwarf (WD) ($1.08$ M$_{\odot}$) and a 4400 K main-sequence companion (0.66 M$_{\odot}$). This WD is among the most massive known within post-common envelope binary systems. We also find, through both spectral energy distribution and $\it{TESS}$ light curve analyses, that the WD has a radius of $1.54 \pm 0.07R_{\earth}$, roughly $12\sigma$ larger than the expected value from WD mass-radius relationships. Both the Lomb-Scargle analysis and the $v \sin{i}$ of the system indicate the main-sequence companion to be super-syncronously rotating at a period of $\sim$6 days, which may suggest accretion occurred during the evolution of the system. This binary also shares similar physical characteristics with six other post-common envelope systems hosting massive WDs, which may point to a shared formation pathway. We model the history of this system with COSMIC and find that it likely formed through an episode of common envelope evolution following the onset of mass transfer when the progenitor primary was on either the early or the thermally pulsing stages of the asymptotic giant branch. As a result of its properties, the study of 2M07515777+1807352 can provide new insights regarding many key outstanding questions in our understanding of common envelope evolution.
\end{abstract}


\section{Introduction}
 Many fundamental questions in astrophysics depend significantly on our understanding of the physical processes involved in binary evolution. Binary systems are progenitors of many astrophysical phenomena of substantial current interest, including cataclysmic variables, X-ray binaries, novae, Type 1a supernovae, and gravitational wave sources \citep{2013A&ARv..21...59I, 2016ApJ...830L..18B, 2019MNRAS.490.5888L, 2024arXiv241104563P}. Stars within binary systems can interact through various mechanisms that dramatically impact their evolution, such as common envelope evolution (CEE), mass transfer (MT) through Roche lobe overflow (RLOF) and wind accretion, and mergers. 

CEE may occur at least once during the evolution of the shortest-period binaries \citep{ 2020RAA....20..135W,2015ApJ...806..135H}, yet the physical processes involved in CEE are among the least constrained of all binary evolution \citep{2013A&ARv..21...59I}. The short-lived nature of CEE events  \citep[400-4000 yr, ][]{1991ApJ...370..709H} make direct observations challenging. Luminous Red Novae have been associated with mass ejections from binary systems undergoing CEE, though they are infrequent and have been measured to only last up to $\sim$$200$ days \citep{2020MNRAS.492.3229H,2003ApJ...582L.105S}. Consequently, observational evidence of CEE relies heavily on the identification of post-CEE binaries (PCEBs).  CEE is also  difficult to model comprehensively due to the complexity of the multiple processes involved, often leading to simulations that make significant simplifications and yield few secure predictions \citep{2013A&ARv..21...59I}. However, with observations and subsequent analysis of likely post-CEE systems, we are able to empirically enhance our understanding of CEE \citep{2025ApJ...979L...1L,2024A&A...686A..61B, 2024MNRAS.52711719Y, 2022MNRAS.517.2867H, 2022MNRAS.512.1843H, 2021MNRAS.501.1677H, 2012MNRAS.423..320R, 2011A&A...536L...3Z, 2010A&A...513L...7S, 2009JPhCS.172a2024S}.  

Current theory suggests that close binaries consisting of a main-sequence (MS) star and a white dwarf (WD) are the products of recent CEE \citep{1976IAUS...73...75P,2013A&ARv..21...59I,1984ApJ...277..355W}. CEE has been thought to arise through two main mechanisms. The first mechanism can begin as the more massive companion (primary) in a close binary evolves and expands, causing it to fill its Roche lobe and begin MT onto the secondary star. If the MT becomes rapid enough to be unstable, both stars can overflow their Roche lobes with a common envelope of material forming around the system. Then, as the secondary plunges toward the primary, the common envelope is ejected, leaving behind a close binary system composed of a newly formed WD (the core of the original primary) and a MS star (the original secondary). Another way a CE can form is through an instability in the tidal interaction (the Darwin instability), causing a spiraling in of the orbit until the secondary plunges into the primary's envelope \citep{2013A&ARv..21...59I}. This is expected for very unequal mass ratios. The subsequent orbital separation in both scenarios will be much less than before the episode of CEE due to the loss of orbital energy used to eject the common envelope \citep{2008MNRAS.390.1635R,2011A&A...536A..43N}. 

Despite recent progress, a comprehensive theoretical understanding of common envelope formation and ejection is still incomplete \citep{2024PrPNP.13404083C, 2020cee..book.....I, 2025RAA....25k5014W, 2025A&A...698A.173T, 2010MNRAS.403..179D}. Particularly relevant here is the question of how much orbital energy is needed to eject the common envelope. Commonly adopted efficiencies lead to PCEBs with orbital periods on the order of a few days or less \citep{2010MNRAS.403..179D}. Since the post-CEE orbital period is thought to be set by the amount of released orbital energy, longer-period systems (greater than a few days) challenge our understanding of the energy required for envelope ejection. The possible importance of additional energy sources, such as recombination energy and accretion energy, has been widely studied \citep{2020cee..book.....I,2013A&ARv..21...59I,2015MNRAS.447.2181I, 2022MNRAS.516.4669L, 2008ASSL..352..233W, 2012MNRAS.423..320R, 2012ApJ...744...52P, 2016MNRAS.460.3992N,2010MNRAS.403..179D, 2024MNRAS.52711719Y,2025RAA....25b5023S,2025RAA....25k5014W}. However, other studies argue that additional energy, beyond orbital energy, is not necessary to form wide binaries following CEE \citep{ 2024A&A...687A..12B, 2019MNRAS.490.2550I}.

Here, we provide an in-depth analysis of 2MASS 2M07515777+1807352 (henceforth, 2M07515777; Gaia DR3 668547707285277440, $\alpha$: 07:51:57.77, $\delta$: +18:07:35.24), a likely PCEB. This system consists of a massive WD (1.08~M$_{\odot}$) in a 10.3-day, nearly circular orbit (e = 0.02) with a MS companion of spectral type K5 (0.66~M$_{\odot}$). 2M07515777 was identified within a subsample of the recently produced APOGEE-GALEX-Gaia Catalog (AGGC; \citealt{2022AJ....164..126A}). The AGGC includes 3414 systematically identified field late-type stars with candidate WD companions (identified through ultraviolet (UV) excesses). The AGGC expands upon previous catalogs through combining The Apache Point Observatory Galactic Evolution Experiment (APOGEE; \citet{2017AJ....154...94M}) high-resolution infrared (IR) spectroscopy, Galaxy Evolution Explorer (GALEX; \citet{2017ApJS..230...24B}) UV photometry, multi-band far-UV/optical/IR spectral energy distributions (SEDs), and Gaia distances \citep{2018yCat.1345....0G}. In addition to the AGGC, this system has also been listed as a candidate WD-MS in other studies \citep{2024MNRAS.529.4840G, 2017MNRAS.472.4193R}. 

Recently, a small sample of other PCEB systems hosting massive WDs ($\gtrsim 1.1$M$_{\odot}$) have been identified \citep{2024A&A...686A..61B,1993PASP..105..841L,2024MNRAS.52711719Y}. Notably, all of the systems in the sample lie in a similar orbital parameter space as 2M07515777, as they have orbital periods that range from 18 to 49 days and very low, but non-zero, eccentricities. 2M07515777 WD is on the boundary of the most massive known WDs in PCEBs. Hence, this binary resides in a relatively unexplored, yet valuable parameter space of PCEB systems that can be leveraged empirically to test currently unconstrained CEE theories regarding energy budgets, envelope ejection, sources of eccentricity, evolutionary stages preceding the onset of MT, and spin angular momentum evolution.

This analysis of 2M07515777 is part of an on-going larger study that aims to conduct multi-variable statistical analyses on and comprehensively characterize an extensive portion of the AGGC, including PCEBs across the Hertzsprung-Russell (HR) Diagram.

In Section \ref{sec:analysis}, we provide a detailed description of the methods we use to characterize the system, including our radial-velocity study, spectral energy distribution fitting, light-curve modeling, and kinematical analysis. We present COSMIC evolutionary models for the system in Section \ref{cosmic}. We then move to a discussion of possible formation pathways and system properties in Section \ref{sec:discussion}, and provide our concluding thoughts in Section \ref{sec:conclusions}.

\section{Data Analysis}\label{sec:analysis}

\subsection{Radial velocities and orbit solution}
\label{subsec:orbit}
We obtain a single-lined spectroscopic orbit solution for the system combining earlier time-series radial-velocity (RV) measurements ($\sigma$ $\approx$ 0.1-0.2 km/s) from APOGEE with additional precise ($\sigma\approx$ 0.1 km/s) RV measurements with the high-resolution (R$~\approx 110,000$) NEID spectrograph at the WIYN 3.5m telescope.\footnote{The WIYN 3.5m Observatory is a joint facility of the University of Wisconsin–Madison, Indiana University, NSF’s NOIRLab, the Pennsylvania State University and Princeton University.} The WIYN/NEID RVs are determined via cross correlation against template spectra using a weighted numerical stellar mask based on spectral type via the NEID Data Reduction Pipeline.\footnote{https://neid.ipac.caltech.edu/docs/NEID-DRP/} The spectra obtained by NEID have signal-to-noise (SNR)
ratios ranging between SNR=0.66-2.53. We include three observations over 289 days obtained from APOGEE and 15 observations over 416 days obtained from NEID for 2M07515777, spanning a total of 2860 days.

\begin{table}[t]
    \setlength{\tabcolsep}{3.5pt}
    \centering
    \caption{Radial-velocity measurements}
    \begin{tabular}{cccc}
    \hline 
    Instrument & JD & RV(km $\text{s}^{-1}$) & Error (km $\text{s}^{-1}$)\\
    \hline  
    APOGEE & 2457783.744& -20.9345 & 0.030\\
    APOGEE & 2458070.989& -46.22768 & 0.074\\
    APOGEE & 2458072.921& 11.97548 & 0.027\\
    NEID & 2460228.015& 97.9722 & 0.041 \\
    NEID & 2460230.933& 14.1892 & 0.042\\
    NEID & 2460234.921& -17.0853 & 0.024\\
    NEID & 2460236.919& 67.7782 & 0.020\\
    NEID & 2460255.030& -32.6499 & 0.042\\
    NEID & 2460327.901& -5.6077 & 0.034\\
    NEID & 2460328.838& 35.7761 & 0.043\\
    NEID & 2460340.885& 94.3052 & 0.046\\
    NEID & 2460341.907& 94.3015 & 0.053\\
    NEID & 2460411.730& 57.4507 & 0.013\\
    NEID & 2460435.630& 67.4860 & 0.020\\
    NEID & 2460436.660& 25.0226 & 0.026\\
    NEID & 2460567.997& 98.0763 & 0.048\\
    NEID & 2460632.914& -0.0197 & 0.042\\
    NEID & 2460643.898& -25.0043 & 0.048\\
    \hline
    \end{tabular}
-\end{table}

\begin{figure}[h!]
    \includegraphics[width=\linewidth]{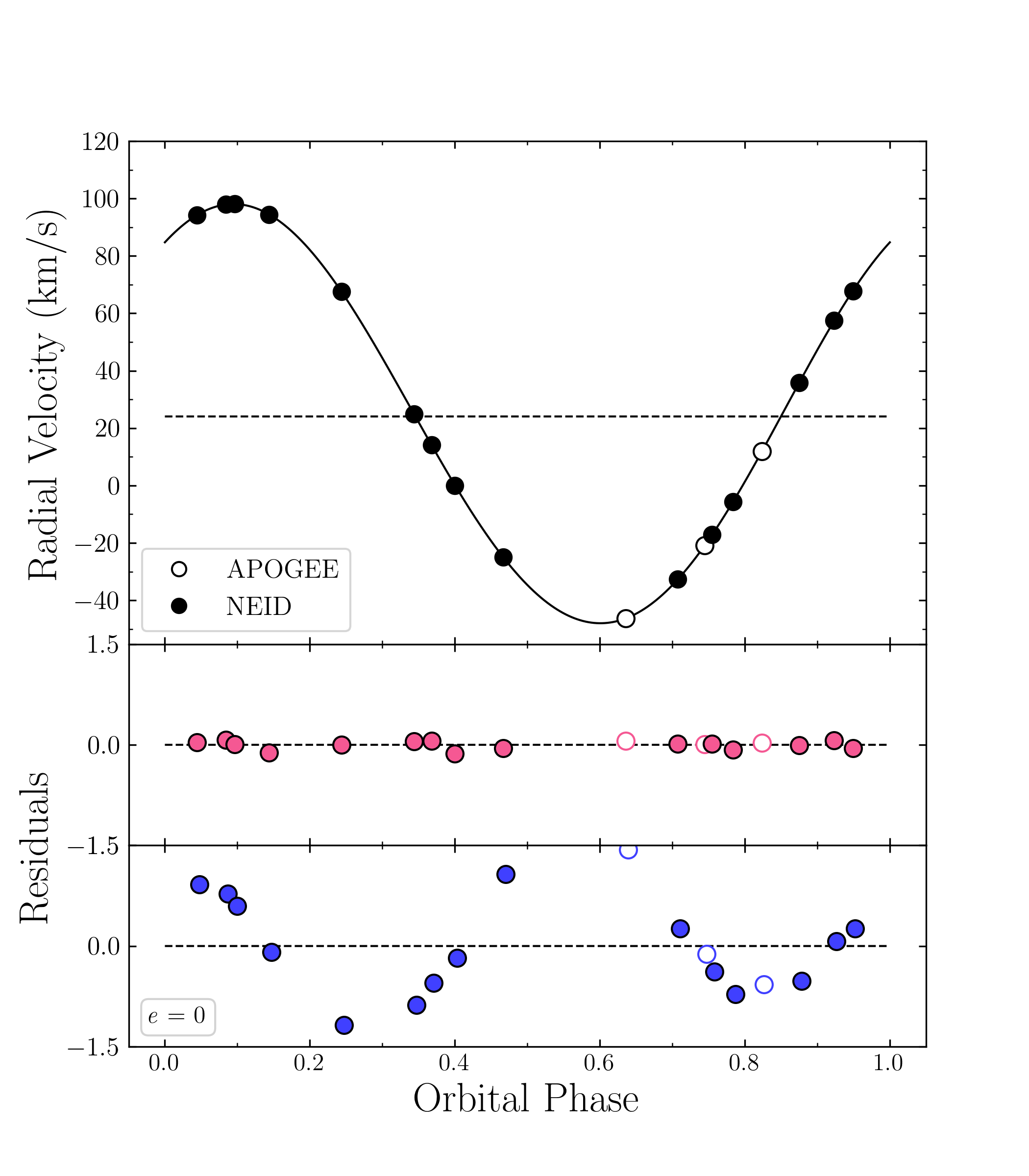}
    \caption{Orbit solution for 2M07515777 (P = 10.3 d, Table \ref{tab:all_param}) as a function of phase (top panel). RV measurements are shown from both APOGEE (open circles) and NEID (filled circles), along with the orbital velocity curve. The center-of-mass velocity, $\gamma$, is shown as a dashed line. In the middle panel, we show the residuals corresponding to the best-fit orbit along with (bottom panel) the residuals of a completely circular (e=0) orbit of the same period to demonstrate the significance of the non-zero eccentricity. The residuals of the best-fit orbit are consistent with the RV measurement errors reported by NEID. 
    \label{fig:fig1}}
\end{figure}

The RV measurements used to determine the orbit are listed in Table 1. The resulting orbit solution is provided in Table \ref{tab:all_param} and shown in Figure \ref{fig:fig1}. The orbit solution is determined with a direct integrator \citep{1973ApJ...186..185C,2002AJ....124.1144L,2002AJ....124.1132G}. The direct integrator has previously established capability to provide secure and precise orbital solutions combining RV measurements from multiple sources \citep{2021AJ....161..190G}. 

We find an orbital period P = 10.29862 $\pm$ 0.000007 days, eccentricity \textit{e} = 0.0164 $\pm$ 0.0004, and semi-amplitude velocity K = 73.03 $\pm$ 0.03 km $s^{-1}$. This orbit agrees well with the published Gaia RV-derived orbit of P = 10.29809 $\pm$ 0.00066 days, eccentricity \textit{e} = 0.022 $\pm$ 0.018, and semi-amplitude velocity K = 72.29 $\pm$ 1.03 km $s^{-1}$\citep{2023A&A...674A..34G}. The errors for the Gaia solution are higher as a result of the lower precision RV measurements compared to NEID. 


Our derived eccentricity and error indicate a nearly, but not quite, circular orbit. Even when formally significant, nearly circular eccentricities are at times thought to be insignificant and a result of systematic biases since negative eccentricities are not possible \citep{1971AJ.....76..544L, 2008ApJ...685..553S}. To investigate the small, yet formally non-zero, measured eccentricity of 2M07515777, we force a circular orbit fit and show the corresponding residuals in the bottom panel of Figure \ref{fig:fig1}, following the methods of \citet{2024MNRAS.52711719Y}. These residuals produced by the \textit{e} = 0 fit are much greater than the best fit orbit solution and indicate that the \textit{e} = 0.0164 value is likely a significant measurement. We discuss this eccentricity further in Section \ref{sec:discussion} through comparisons with similar systems in the literature.

We obtain the mass of the MS companion from the APOGEE database which was determined using \texttt{starHorse} with APOGEE measurements (see Table \ref{tab:all_param}). The \texttt{starHorse} algorithm is a Bayesian tool used for determining stellar masses, ages,
distances, and extinctions for field stars. The software employs a method that combines spectroscopically measured stellar properties such as effective temperature, surface gravity, and metallicity (which for this paper, are taken from the APOGEE survey) along with photometric magnitudes and parallax \citep{2018MNRAS.476.2556Q}. The reader is directed to \citet{2018MNRAS.476.2556Q} for a detailed description of the software and its methodology. These parameters are then compared to stellar evolutionary models using a statistical approach. Combining the resulting K-dwarf mass estimate (M=$0.66\pm0.08$ M$_{\odot}$; \citet{2020A&A...638A..76Q}) with the mass function from the orbital solution, we find a lower limit for the companion mass of  $1.085 \pm0.06$ M$_{\odot}$.

\subsection{Spectral Energy Distribution}
\label{subsec:sed}
\begin{figure*}[ht!]
\centering
\includegraphics[width=.9\linewidth,trim=80 50 40 50,clip]{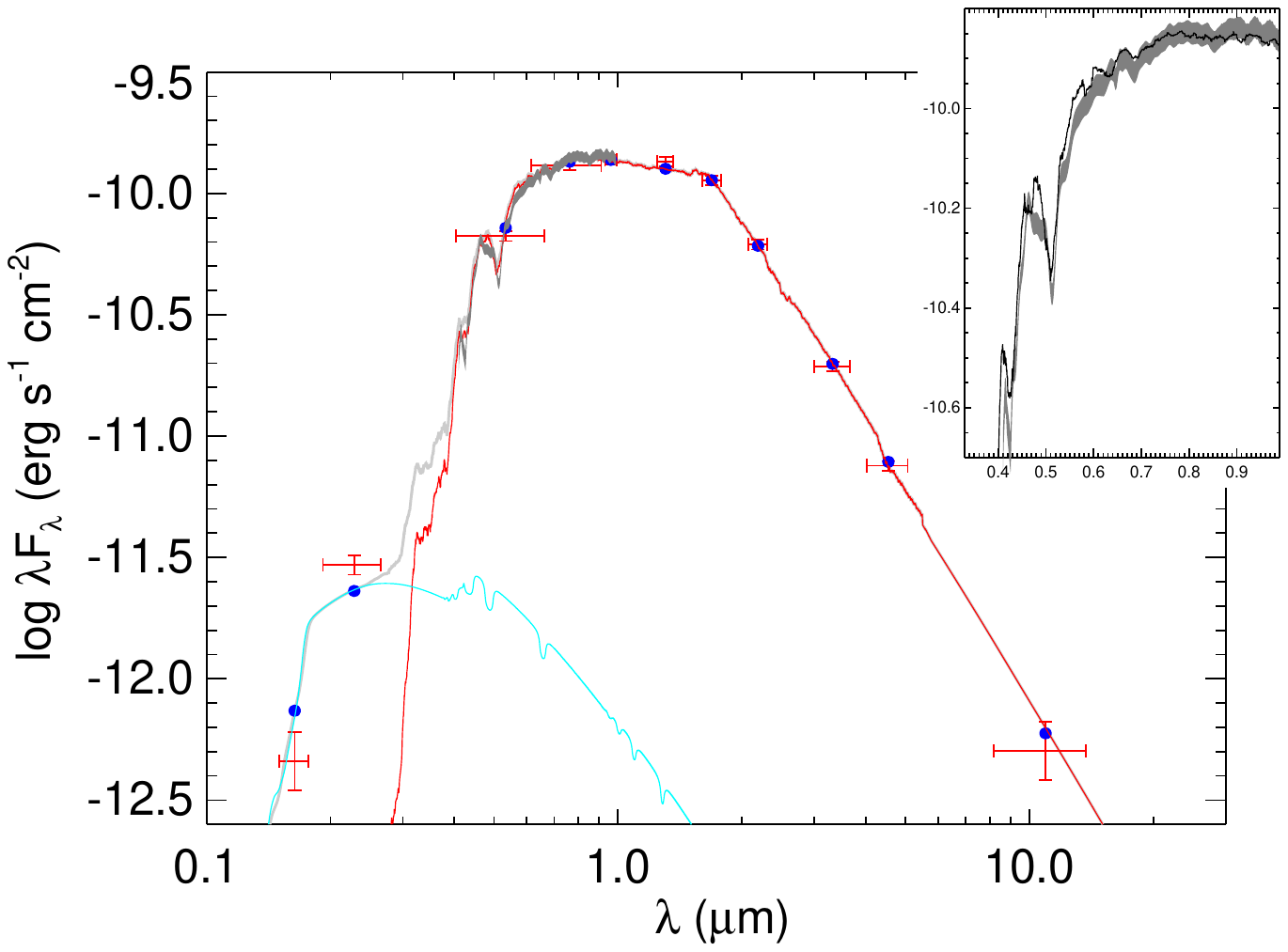}
    \caption{Spectral energy distribution of 2M07515777. Red symbols represent the observed photometric measurements, where the horizontal bars represent the effective width of the passband. Blue symbols are the model fluxes from the best-fit 2-component atmosphere model (red curve for the K dwarf, cyan curve for the WD, grey curve for the two combined). The inset shows the absolute flux-calibrated {\it Gaia\/} spectrophotometry as a grey swath overlaid on the 2-component best-fit model. \label{fig:sed}}
\end{figure*}

\begin{figure*}[ht!]
\centering
\includegraphics[width=\linewidth]{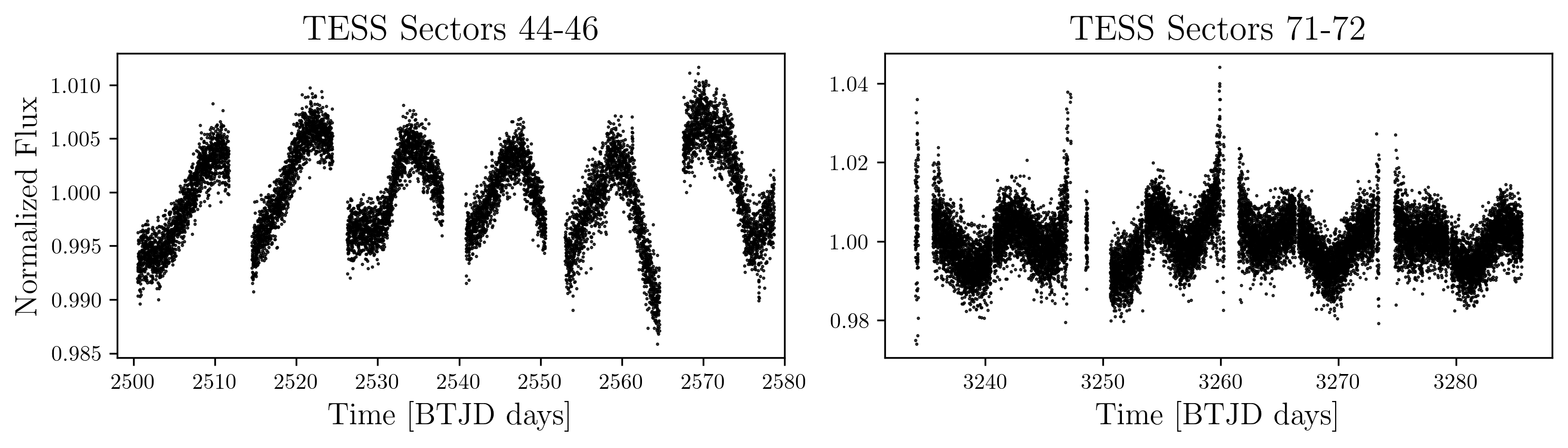}
    \caption{Detrended \textit{TESS} light curves for 2M07515777, including sectors 44-46 (left) and 71-72 (right), both showing periodic variation.  
    \label{fig:lc}}
\end{figure*}

To establish the presence of a hot WD companion and to determine its physical parameters, we analyzed the broadband SED of the system together with the {\it Gaia}  DR3 parallax \citep[with no systematic offset applied; see, e.g.,][]{StassunTorres:2021}, following the procedures described in \citet{Stassun:2016,Stassun:2017,Stassun:2018}. We obtained the $JHK_S$ magnitudes from 2MASS \citep{2003yCat.2246....0C}, the W1--W3 magnitudes from WISE \citep{2019ApJS..240...30S}, the $G_{\rm BP}\mathrm{\;and\;} G_{\rm RP}$ magnitudes from Gaia \citep{2022yCat.1355....0G}, the $y$ magnitude from Pan-STARRS \citep{2016arXiv161205560C}, and the FUV and NUV magnitudes from GALEX \citep{2017ApJS..230...24B}. We also utilized the absolute flux-calibrated {\it Gaia\/} spectrophotometry. Together, the available photometry spans the full stellar SED over the wavelength range 0.2--10~$\mu$m (Figure~\ref{fig:sed}).

The SED provides several lines of evidence suggesting that the 1.1 M$_{\odot}$ companion is a WD. At a Gaia distance of 137 pc, the observed optical fluxes of the system are several orders of magnitude less than expected from a 1.1 M$_{\odot}$ MS star. (The single-star photometry is also consistent with the single-lined K-dwarf NEID spectrum.) The maximum flux of the SED is not consistent with the temperature of a 1.1 M$_{\odot}$ MS star and qualitatively the GALEX UV flux measurements indicate UV excess flux. 

As a quantitative analysis, we computed a two-component fit using PHOENIX stellar atmosphere models \citep{Husser:2013} for the MS primary star and \cite{2010MmSAI..81..921K} atmosphere models for a WD companion. For the K dwarf, the effective temperature ($T_{\rm eff}$ = $4400 \pm 10$ K) and metallicity ([Fe/H] = $-0.09 \pm 0.01$) were adopted from the APOGEE spectroscopic measurements \citep{2023yCat.3286....0A}. We also imposed parameter constraints using the measurements and uncertainties of WD radius and temperature ratio from the PHOEBE eclipse analysis (see Section \ref{subsec:lightcurve}). The fitted parameters were the WD radius and temperature, as well as the extinction to the system, $A_V$, which we limited to the maximum line-of-sight value from the Galactic dust maps of \citet{Schlegel:1998}. 

We were able to achieve a goodness-of-fit of $\chi_\nu^2 = 1.9$ with $A_V = 0.09 \pm 0.04$, $T_{\rm WD} = 10,900 \pm 600$~K, and $R_{\rm WD} = 1.54 \pm 0.07$~R$_\oplus$. The WD temperature and radius found in this analysis is within 1.1$\sigma$ of the values determined from the eclipse analysis (T$_{\rm eff}$=$8429_{-1130}^{+2122}$ K and $R$=$1.30_{-0.25}^{+0.31}$$R_{\oplus}$; Table \ref{tab:all_param}). This solution is also shown in Figure~\ref{fig:sed}, both for each component and as the combined light.

We note for completeness that it is possible to achieve an SED fit with reduced $\chi^2 = 1.0$ with a slightly cooler WD temperature (within 1.0$\sigma$ of the eclipse constraint) and a somewhat larger WD radius (nearly $3\sigma$ larger than the eclipse value). We have opted for the above values as representing an empirically good fit while also best matching the constraints from the eclipse analysis. 

\subsection{Light curve analyses}\label{subsec:lightcurve}

We use \textit{Transiting Exoplanet Survey Satellite} \citep[\textit{TESS}; ][]{2022BAAS...54e4602R})
Simple Aperture Photometry (SAP) flux data for the system, including sectors 44-46 (600s cadence) and 71-72 (200s cadence), all shown in Figure \ref{fig:lc}. We then subsequently remove any outliers and systematic trends within the data using the \texttt{keplersplinev2} package and sigma-clipping.

Analysis of the light curve for the system reveals that this binary is eclipsing and specifically that the WD is being occulted by its MS companion (see Table \ref{tab:all_param} for eclipse parameters). By modeling the eclipse we obtain a direct measurement of the WD radius along with additional measurements of the other orbital and stellar properties of the system.  

\subsubsection{Modeling with PHOEBE}
We model the occultation of the WD with the PHysics Of Eclipsing BinariEs (\textsc{phoebe} software; \citep{2020ApJS..250...34C}). \textsc{phoebe} is an eclipsing binary modeling code that can we used to reproduce and fit light curves of eclipsing systems. 

We utilize the Nelder-Mead algorithm in \textsc{phoebe} to optimize the orbital period, inclination, time of super-conjunction, the sum of the equivalent fractional radii of two stars in a binary system\footnote{requivsumfrac}, and WD temperature. We choose to keep the properties of the MS star fixed, as they are well defined from APOGEE measurements and SED fitting. These fixed parameters are $R_{ms}$, $M_{ms}$, and $T_{ms}$. We also set the WD mass to remain constant as it is highly constrained by the mass function in combination with the inclination ($i \approx 90$ deg) from the eclipse. 

Additionally, we set PHOENIX and blackbody models to be used for the MS star and WD, respectively. We also use typical values for bolometric gravity-darkening coefficients ($\beta_{wd} = 1.0$ and $\beta_{ms} = 0.32$) \citep{1924MNRAS..84..702V,1967ZA.....65...89L}. To account for irradiation effects, the reflection fraction for the MS star is set to 50\% and the WD is again treated as a blackbody. To approximate the effects of limb-darkening for our WD component, we adopt logarithmic law coefficients (e=0.424 and f=0.288) from \citet{2020yCat..36410157C}, which were calculate near our WD SED fitted $T_{\rm eff}$ for the TESS filter. The limb-darkening for our MS component is determined automatically for each generated model using \textsc{phoebe}'s built-in atmospheric table interpolation. 

\begin{figure}[h!]
  \includegraphics[width=1.06\linewidth]{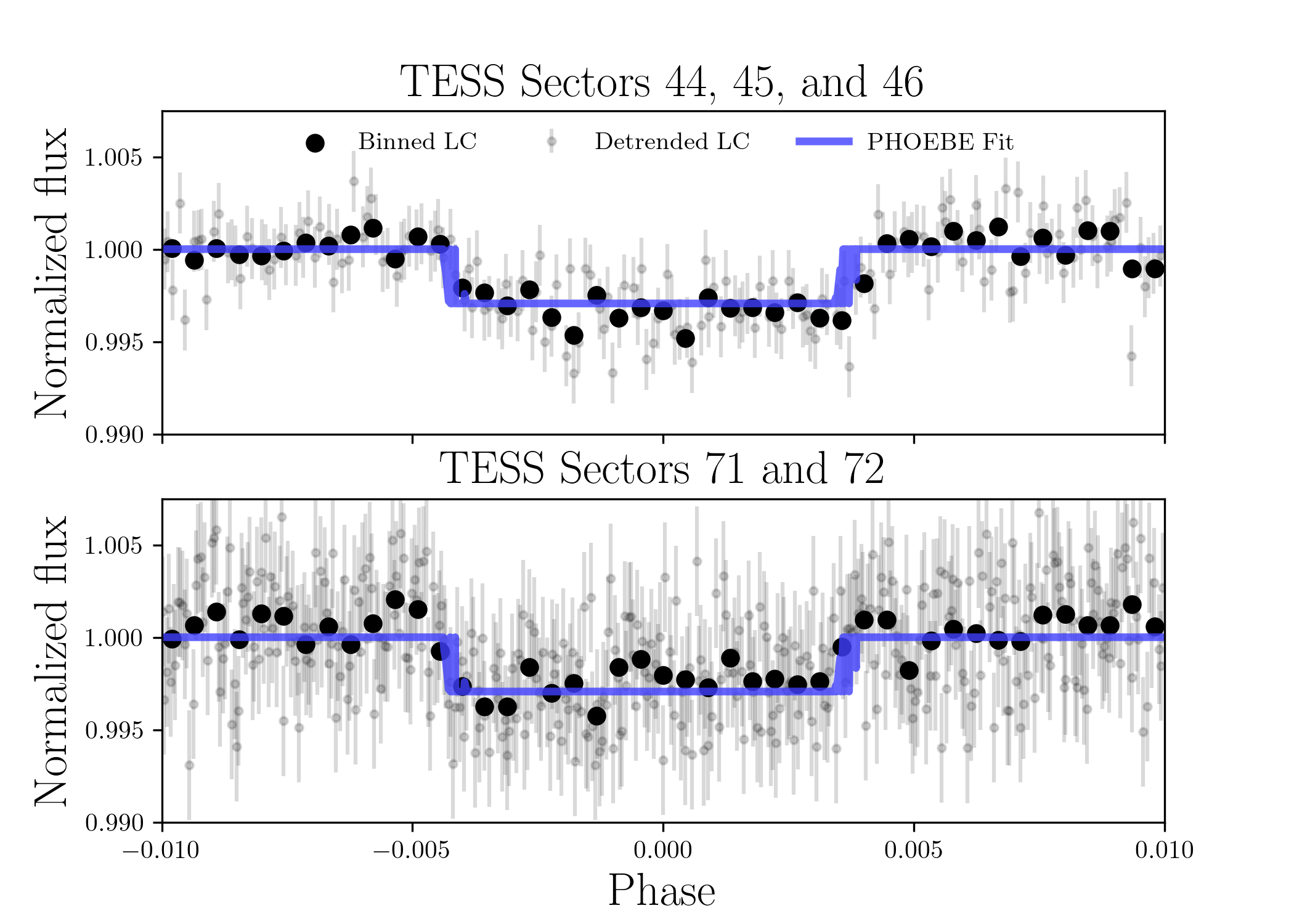}
    \caption{The detrended phase-folded \textit{TESS} light curves of the eclipse and associated errors for 2M07515777 are shown in gray, including sectors 44-46 (upper panel) and 71-72 (lower panel). We also show the binned data with black markers. Overplotted in pink is the resulting fit from PHOEBE using the Nelder-Mead optimizer.  
    \label{fig:fig2}}
\end{figure}

The resulting physical parameters from \textsc{phoebe} are listed in Table \ref{tab:all_param}, with the computed light curve model shown in Figure \ref{fig:fig2}.

Parameter uncertainties were estimated in \textsc{phoebe} using the Markov chain Monte Carlo (MCMC) sampler package  \texttt{emcee}. We used the final values produced from the Nelder-Mead optimization as centers of the initial gaussian sampling distributions. To eliminate degeneracies in the fit, we also impose a prior with a maximum WD temperature of 13800 K, which was determined by allowing the SED fit to deviate from the NUV and FUV GALEX photometry by $3.5 \sigma$, respectively. The $1 \sigma$ confidence intervals produced by \texttt{emcee} for all fitted parameters are shown in Table \ref{tab:all_param}.

\subsubsection{Super-synchronous Rotation
}
\label{subsubsec:rot}

 A notable property of 2M07515777 is the rotation of the K-star companion. The projected rotational velocity and radius measurements (from APOGEE/\texttt{starhorse}) are $v \sin{i} = 6.3$ km s$^{-1}$ and $R_{ms} = 0.67 R_{\odot} \pm 0.07$, respectively. These values yield a rotational period of $5.4 \pm 0.5$ days---approximately half of the 10.3 day orbital period for the system. We do note that APOGEE does not report uncertainties on $v \sin{i}$, which could lead to an underestimate of the rotational period uncertainty.

\begin{figure}[h!]
    \includegraphics[width=\linewidth]{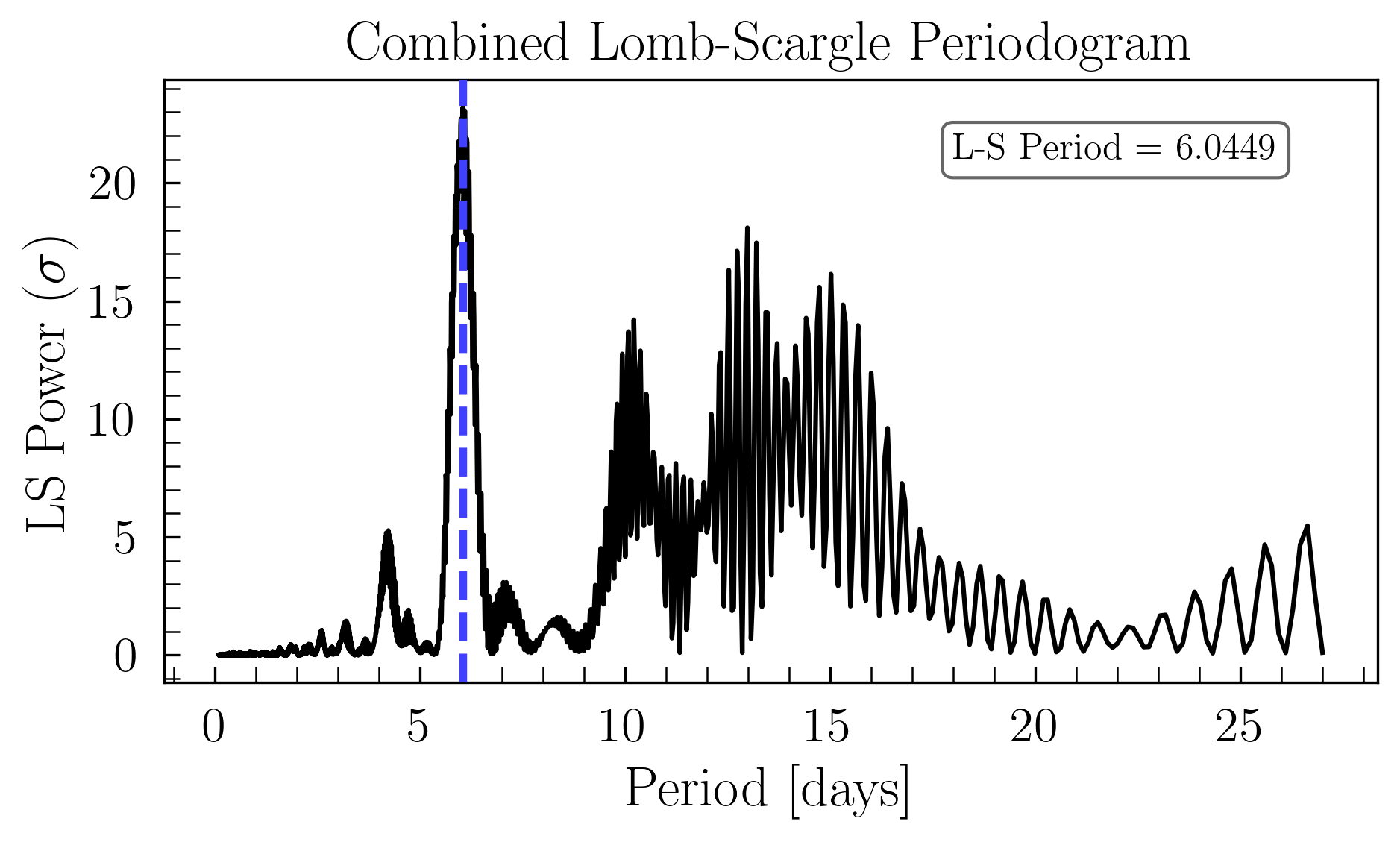}
    \caption{Lomb-Scargle periodogram of all combined \textit{TESS} light curve sectors, indicating a peak period of 6.0449 days.
    \label{fig:ls}}
\end{figure}

The TESS light curves show periodic variation (Figure \ref{fig:lc}). We apply a Lomb-Scargle \citep{1976Ap&SS..39..447L,1982ApJ...263..835S} analysis on the combined light curve (sectors 44-46 and 71-72). As seen in Figure \ref{fig:ls}, we identify a strong peak at $6.0429 \pm 0.0002$ days. This value is approximately 1$\sigma$ from the period of 5.4 days derived through $v \sin{i}$. However, both values suggest that the MS companion is rotating super-synchronously at approximately 6 days.  

Additionally, we conduct a Lomb-Scargle analysis on each sector individually and find a recurrent peak of $\sim$12 days for sectors 44-46 as well as $\sim$6 and $\sim$11 day peaks for sectors 71 and 72, respectively. We suspect that these peaks are harmonics, but with the currently available data alone, we are unable to determine the true peaks of the individual sectors confidently.

\begin{table*}[ht]
\centering
    \setlength{\tabcolsep}{15pt}
    \renewcommand{\arraystretch}{1.25}
    \caption{ Summary of Binary Properties }
    \begin{tabular}{ccc}
    \hline 
    \textbf{Parameter} &  \textbf{Value} & \textbf{Method} \\
    \hline
    \hline
    \multicolumn{3}{c}{\textbf{Orbital Parameters}} \\
    \hline
    P (days) & $10.29862 \pm 0.00001 $ &RV \\
    $\it{e}$ & $0.01643 \pm 0.00040$ &RV \\
    $\gamma$ (km $s^{-1}$)& $24.06 \pm 0.02$ &RV \\
    K (km $s^{-1}$)& $73.03 \pm 0.03$ &RV \\
    $\omega$ (deg) & $325.00 \pm$ 1.12 & RV \\
    $T_0$ (BJD - 2400000) & $59979.98 \pm 0.03$ &RV \\
    a sin$i$($10^6$ km)& $10.341 \pm 0.004$ &RV \\
    f(\it{M}) ($M_{\odot}$) & $0.415 \pm 0.001$ &RV \\
    $\sigma$ (km $s^{-1}$) & 0.071 & RV \\
     \hline
    \multicolumn{3}{c}{\textbf{Eclipse Parameters}} \\
    \hline
    P (days)& $10.29856 \pm 0.00003$ &\textsc{phoebe}\\
    $\it{i}$ (deg) & $89.25 \pm 0.05$ &\textsc{phoebe}\\
    $t_0$ (BJD - 2457000) &  $3266.8142_{-0.0010}^{+0.0009}$ &\textsc{phoebe}\\
    Eclipse Duration (hrs)& $2.616$ &\textsc{TESS}\\
    Eclipse Depth (ppm)& $4218.7$ &\textsc{TESS}\\
    \hline
    \multicolumn{3}{c}{\textbf{White Dwarf Parameters}} \\
    \hline
    M ($M_{\odot}$) &  $ 1.08$ $\pm$ $0.06$ & RV \\
    R ($R_{\oplus}$)& $1.30_{-0.25}^{+0.31}$ &\textsc{phoebe}\\
    T$_{\rm eff}$(K) & $8429_{-1130}^{+2122}$ &\textsc{phoebe}\\
    R ($R_{\oplus}$)& $1.54 \pm 0.07$  &SED\\
    T$_{\rm eff} $(K) & $10,900 \pm 600$ &SED\\
    \hline
    \multicolumn{3}{c}{\textbf{Main Sequence Star Parameters}} \\
    \hline
    M ($M_{\odot}$) & $0.66 \pm 0.08$ &APOGEE/\texttt{starHorse} \\
    R ($R_{\odot}$)& $0.67 \pm 0.07$ &APOGEE/\texttt{starHorse} \\
    $log(g)$ ($cm/s^2$) & $4.60 \pm 0.02$ & APOGEE\\
    T$_{\rm eff}$(K) &  $4400 \pm 10$ &APOGEE\\
    \hline
    \hline
    \end{tabular}
    \label{tab:all_param}
\end{table*}
\subsection{Speckle Imaging}

2M07515777 was observed with the Differential Speckle Survey Instrument (DSSI; \citealt{Horch2009}) on the Astrophysical Research Consortium (ARC) 3.5-meter telescope at Apache Point Observatory (APO) \citep{Davidson2024} on both UT 24 Sep 2024 and 15 Nov 2024.  By use of a dichroic to split the beam, DSSI takes speckle observations simultaneously in two passbands; in this case images were taken in passbands with central wavelengths of 692 and 880nm with wavelength widths of 40 and 50nm, respectively.  Both nights on which the source was observed started out with excellent conditions (e.g., seeing of $\sim$0.9 and 0.5 arcseconds, respectively), but due to the source coordinates it had to be observed at the end of the night, and, due to the source faintness, it was given especially long net integration time.  The nominal DSSI speckle image exposures are 40 msec, and a sequence of 9,000 such exposures were obtained on UT 15 Nov 2024.  However, on UT 24 Sep 2024 conditions were such that it was deemed necessary to increase the frame exposures to 100 msec for the 9,000 frame sequence obtained that evening.  In each case, a 1000-frame observation with the same individual frame exposure length of a nearby reference point source, HR 3086, was obtained just after observations of 2M07515777.

Reductions of these speckle data followed the standard pipeline historically followed for DSSI data as described in \citet{Horch2011a}, \citet{Horch2021}, and (as specifically applied for DSSI on the ARC 3.5-m telescope) in \citet{Davidson2024}.  First a power spectrum and bispectrum of the observations are computed from the speckle frames.  Then a reconstructed image—deconvolved by the paired point source observation and low-pass filtered to reduce noise—is formed by assembling the diffraction-limited Fourier transform of the source from the spatial frequency spectra.  For those targets for which a second source is identified in the reconstructed image, a weighted least-squares fit to fringes in the object's power spectrum is computed to obtain the separation, position angle, and magnitude difference of the secondary star relative to the primary. 

However, in the case of both sets of observations of 2M07515777, no resolved secondary source was detected in the observations in either wavelength channel.  In this case, contrast limits on the non-detection are determined as follows:  For each filter, pixel values within 0.1 arcsec-wide circular annuli centered on the primary star are used to determine the statistical variation in the background of the reconstructed image, and the $5\sigma$ value is used to set the $\Delta m$ contrast limit ruling out the presence of a secondary source in the field.  Using the most restrictive results from the combined output of the two observing runs, we can safely rule out the presence of a resolved secondary source at $\lambda=692$ nm at $\Delta m \sim 4.0$ from 0.15 to 1.2 arcsec (which is 21-164 AU for a distance of 137 pc), while for $\lambda = 880$ nm we can safely rule out the presence of a secondary source at $\Delta m \sim 3.0$ from 0.4 to 1.2 arcsec (55-164 AU), dropping down linearly to $\Delta m \sim 3.0$ from 0.4 to 0.1 arcsec (55 to 21 AU).  Note that \cite{Davidson2024} report that DSSI delivers a photometric precision of $0.14 \pm 0.04$ magnitudes when used with the ARC 3.5-m telescope.

Thus, the speckle imaging confirms that there is no third, long-period ($>72$ years) stellar source, at least within a few magnitudes of brightness to the current main sequence primary, that may be complicating the system or our analysis.

\subsection{Possible Confounding Scenarios}

Below, we discuss several scenarios that could lead to a mistaken identification of a WD or impact our measurements of certain characteristics of the system, such as temperature and radii. 

\subsubsection{Chromospheric Activity on the Primary Star}

One major source of potential contamination for this system is UV flux from chromospheric activity caused by low-mass MS stars (i.e., late-K or M stars) in the system.

The 4400 K temperature and $0.66 M_\odot$ mass of the primary star indicates a mid-to-late K-type star. Traditional markers of substantial chromospheric activity in late-type dwarfs include strong emission in H$\alpha$, the Ca H \& K lines, and the Ca II IR triplet \citep{2020AJ....160..269M}. Spectroscopic analyses of H$\alpha$ from stacked NEID spectral observations and from the ARC 3.5m Dual Imaging Spectrograph (taken 2024 January 13) show no evidence of H$\alpha$ emission. Although the Ca H \& K lines in the NEID spectra are too weak to provide meaningful constraints on activity levels, all three lines of the Ca II triplet show deep absorption features with only slight emission cores. In addition, the light-curves shown in Figure \ref{fig:lc}  display periodic variation between sectors which could be a result of spot modulation on the active primary star. Together, these features indicate the presence of at most low-level chromospheric activity on the K-dwarf. 

To assess the potential contribution from chromospheric activity, we use NUV chromospheric excess values for comparable active K-dwarfs calculated by \citet{2016MNRAS.463.1844S} to determine the maximum contribution of NUV flux from the primary's chromosphere. We find that the maximum amount of flux which can be contributed by the K-dwarf's chromosphere is $\sim$$4\times10^{-13}$ $ergs$ $s^{-1} cm^{-2}$, which can account for roughly $90 \%$ of the flux difference between the NUV model and observed NUV photometry in Figure \ref{fig:sed}. However, this flux contribution is still many orders of magnitude too small to explain our measured WD radius of $ 1.54~R_{\oplus}$

Additionally, during our SED fitting (see Section \ref{subsec:sed}), we fit several active early-M dwarf star spectra from the MUSCLES Survey \citep{MUSCLES}.\footnote{\url{https://archive.stsci.edu/doi/resolve/resolve.html?doi=10.17909/T9DG6F}}. The far UV flux of these stars was also several orders of magnitude too low to explain the observed UV flux. 

Thus we conclude that the activity on the primary is not sufficient to contribute enough UV excess to account for our observations of an over-sized WD.

\subsubsection{Chromospheric Activity on a Companion}

A possible source of confusion is chromospheric activity of a MS companion to the K dwarf. An example of this are BY Draconis stars---short-period binaries with a very rapidly rotating K or M star companion that produce chromospheric UV excesses with similar FUV-NUV colors to those of cool white dwarf stars. If such a star were rotating fast enough to produce very broadened lines, we may not detect it in the NEID spectra. We also note that \citet{2024MNRAS.529.4840G} removed 2M07515777 from their WD-MS sample, labeling the system as a contaminant. This work used many conservative cuts to eliminate WD-MS false-positives; it is unclear which flag this system failed. We suspect that this system was cut because they were able to fit a low-mass MS companion in the IR. 

The strongest evidence that the 2M07515777 system is a WD-MS system and not a MS-MS binary is that there is a stellar-mass planet-sized object eclipsing the primary star with the same period as the primary single-lined orbit solution. 

Furthermore, in addition to the speckle non-detection, the very small and non-periodic residuals in the orbital solution suggest that there is not a third star in the system with a period $<$10,000 days, as that would produce another periodic signal in the RV measurements.

\subsubsection{Accretion Disk}

Accreting WDs can produce significant NUV emission. They also are often associated with x-ray emission originating from processes on or near the WD \citep{2017PASP..129f2001M}. 2M07515777 was serendipitously imaged by XMM-Newton on 2000 October 01 (\cite{2020A&A...641A.136W}; 4XMM J075157.6+180734). XMM reported 34 $\pm$ 8 x-ray counts yielding low flux measurements with large errors, consistent with either noise or a K-dwarf just above background levels (private communication M. Orio). 

This non-detection of x-rays alone cannot disprove the presence of an accretion disk around the WD, since accretion disks can radiate in the UV and optical without associated X-ray emission. Another diagnostic of accretion disks are prominent spectroscopic emission features \citep{2017PASP..129f2001M}. The optical spectra of accreting WD systems, particularly those viewed at high inclination \citep{2015MNRAS.450.3331M}, are often characterized by strong single- or double-peaked Balmer emission lines including H$\alpha$, H$\beta$, and H$\gamma$ \citep{2025ApJ...986...34Z}. Since 2M07515777 is an eclipsing system, any accretion disk would be viewed at high inclination and should therefore produce easily detectable Balmer emission. However, as noted above, NEID spectroscopic observations of H$\alpha$, H$\beta$, and H$\gamma$ do not show any strong emission features. This lack of line emission strongly disfavors the presence of an accretion disk and thus suggests that the observed UV excess of this system is not produced by disk accretion.

\subsection{Association with the Praesepe Cluster}
Praesepe (M44 or NGC 2632) is a well studied, nearby open cluster that is in proximity to 2M07515777 (11.5 degrees in projection from cluster center, 3.6 degrees from nearest detected tidal tail member). 2M07515777 was also listed as a potential cluster member by \citet{2023ApJ...954..134K}. Membership would provide valuable information on the age, stellar properties, and potential formation mechanisms of the binary.  We investigate the possibility that 2M07515777 is an ejected cluster member through kinematical analysis, relevant timescales, and a comparison of cluster and stellar properties.

\begin{figure*}[ht!]
\centering
\includegraphics[width=\linewidth]{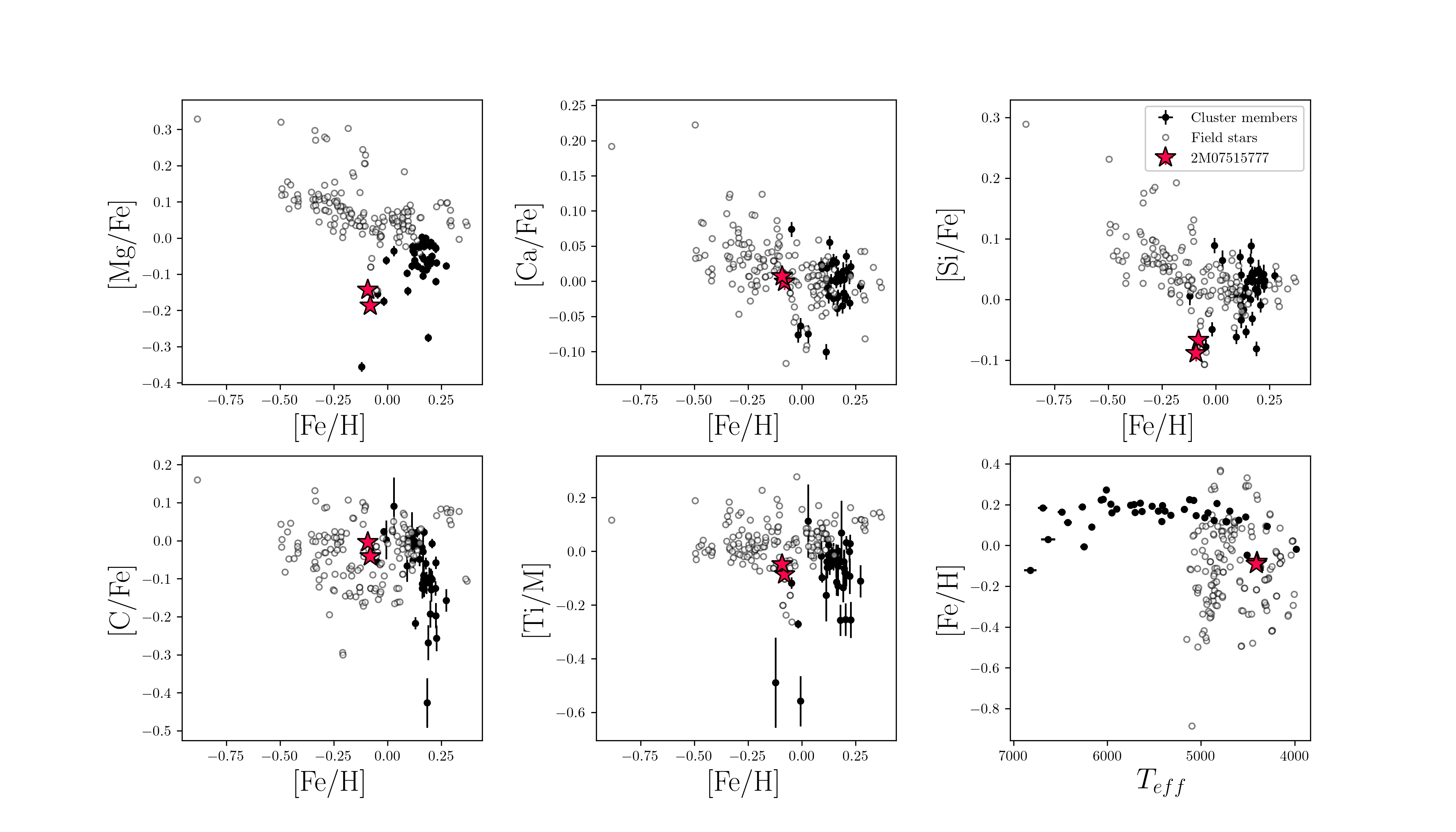}
    \caption{APOGEE abundances for Praesepe cluster members (filled circles), field stars (open circles), and 2M07515777 (red stars; two measurements).  \label{fig:abun}}
\end{figure*}

\subsubsection{Kinematics}
We first utilize precise RV, distance, and proper-motion (PM) measurements from Gaia DR3 \citep{2020yCat.1350....0G} to trace linear paths back in time for both the cluster and the binary. We performed $10^3$ bootstrap iterations, from which we determine the closest approach of 2M07515777 to the cluster core to be $20.0 \pm 0.9$ pc approximately $3$ Myr ago.

Due to the short amount of time since closest approach, it is unlikely that the trajectories were altered by gravitational effects. \citet{2025arXiv250406252H} find that for high-velocity objects ejected from an open cluster, the impacts of the cluster and Galactic gravitational potentials are insignificant. 2M07515777 has a total velocity of 40 km s$^{-1}$ and a relative three-dimensional velocity to the cluster of 17 km/s, to be compared to internal velocities in the cluster of less than 1 km/s. Therefore, we neglect the effects of cluster and Galactic potentials during trajectory calculations.

Though it is possible to have an ejection from the halo of a cluster, it is expected that most runaway ejections ($> 30$ km s$^{-1}$ total velocity) originate from dynamical interactions within $\sim$1 pc from the cluster center \citep{2022A&A...663A..39B}.
Considering that the tidal radius for the cluster is $\approx$ 11 pc, as determined by \citet{2007AJ....134.2340K} and \citet{2019A&A...627A...4R}, and that the closest approach is calculated to be near double this value, we conclude that 2M07515777 was not dynamically ejected from the cluster core or from the bound cluster stars.

\subsubsection{Timescales}\label{subsubsec:timescale}

The cooling time age, $\tau_{cool}$, for a 1.1 $M_{\odot}$ WD with a temperature of $T_{wd}\approx 11000$K is 1.4 $\pm$ 0.070 Gyr \citep{2024MNRAS.529.4840G,2020ApJ...901...93B}, which compared to the cluster age of $\approx 600 $ Myr \citep{2019A&A...628A..66L,2018ApJ...863...67G,2011ASPC..448..841D} also suggests that the system is not an ejected member. We do note, however, that this theoretical cooling time may not hold true for this WD due to the implications of its unusually large radius (further discussed in Section \ref{sec:discussion}). 

\subsubsection{Abundances}

Finally, we analyze and compare the available abundance measurements for the K-dwarf and Praesepe cluster members \citep{2019yCat..36270004R}, all from APOGEE. We also include a sample of field stars within $\sim1^{\circ}$ of 2M07515777. These abundances are shown in Figure \ref{fig:abun}. 

 Previously measured values of metallicty for Praesepe range from [Fe/H] = +0.04 to [Fe/H] = +0.27 \citep{2013ApJ...775...58B,2019A&A...628A..66L}. We find mean metallicities for the cluster and for 2M07515777 of [Fe/H] = $+0.15 \pm 0.07$ (in agreement with the literature) and [Fe/H] = $-0.09 \pm 0.01$, respectively. Thus, the K-dwarf is significantly lower in [Fe/H] to the cluster ($3\sigma$ from the mean of the cluster distribution). 
 
 However, the minimum metallicity measurement within the cluster distribution is [Fe/H] = -0.12, which is inclusive of the value we determined for 2M07515777. It should be noted that the cluster members that have similar metallicities to 2M07515777 are in temperature regimes where less reliable abundance measurements are expected. Specifically, the APOGEE team warns that stars between $4000$ K $<T_{\rm eff}<5000$ K may have abnormally low APOGEE abundance measurements, possibly due to strong lines.\footnote{\url{https://www.sdss4.org/dr17/irspec/abundances/}} Finally, they note specifically that dwarfs cooler than $\sim$4500 K  might show systematic temperature trends for several elements (Holtzman et al. (in prep.)). These trends might explain the scatter for both temperature extremes in the bottom right panel of Figure \ref{fig:abun} as well as the lower metallicity measurements for 2M07515777. 

We also consider several $\alpha$-elements, including C, Mg, Si, Ca, and Ti. We find that the C, Ca, and Ti abundance measurements for 2M07515777 are consistent with the cluster distribution, at $0.6\sigma$, $0.2\sigma$, and $0.1\sigma$ from the cluster average, respectively. Mg and Si deviate more from the cluster, at $1.4\sigma$ and $2.2\sigma$ from the mean. None securely rule out association with the cluster.

On the basis of abundances alone, we cannot securely exclude 2M07515777 as having been a  member of Praesepe. However, combining kinematics, timescales, and abundances, we conclude that 2M07515777 is highly unlikely to be an ejected member of Praesepe.

\section{Modeling with COSMIC} 
\label{cosmic}
We seek to obtain insight into possible formation scenarios for 2M07515777 and the parameters of its progenitor binary with the Compact Object Synthesis and Monte Carlo Investigation Code (COSMIC; \citet{Breivik2020}). COSMIC is a rapid binary population synthesis suite that can be used to simulate the evolutionary stages and orbital characteristics of a binary.

Using COSMIC, we evolved a grid of binaries encompassing evenly spaced ranges of initial parameters including the following: initial orbital periods between 100-7000 days (5000 increments), progenitor primary masses between 4.5-8 $M_{\odot}$ (30 increments) and progenitor secondary masses between 0.6-0.8 $M_{\odot}$ (10 increments). The selected range of initial orbital periods is informed by exploratory COSMIC runs and the small range of progenitor secondary mass is included to account for the uncertainty of the MS star mass measurement and possible mass accretion onto the secondary. The range of progenitor primary mass reflects the full extent of possible initial masses within current WD initial-final mass relations which we calculated as follows. 

Using our measurement for the current WD mass, we use both PARSEC \citep{2012MNRAS.427..127B} and MIST-based \citep{2016ApJ...823..102C, 2016ApJS..222....8D} initial-final mass relations to estimate a WD progenitor mass following \citet{2018ApJ...866...21C}. We calculate progenitor primary masses of $M =6.24 \pm 1.56 M_{\odot}$ and $M =5.75 \pm 1.25 M_{\odot}$ for PARSEC and MIST models, respectively.
\begin{figure*}[ht!]
\centering
\includegraphics[width=\linewidth]{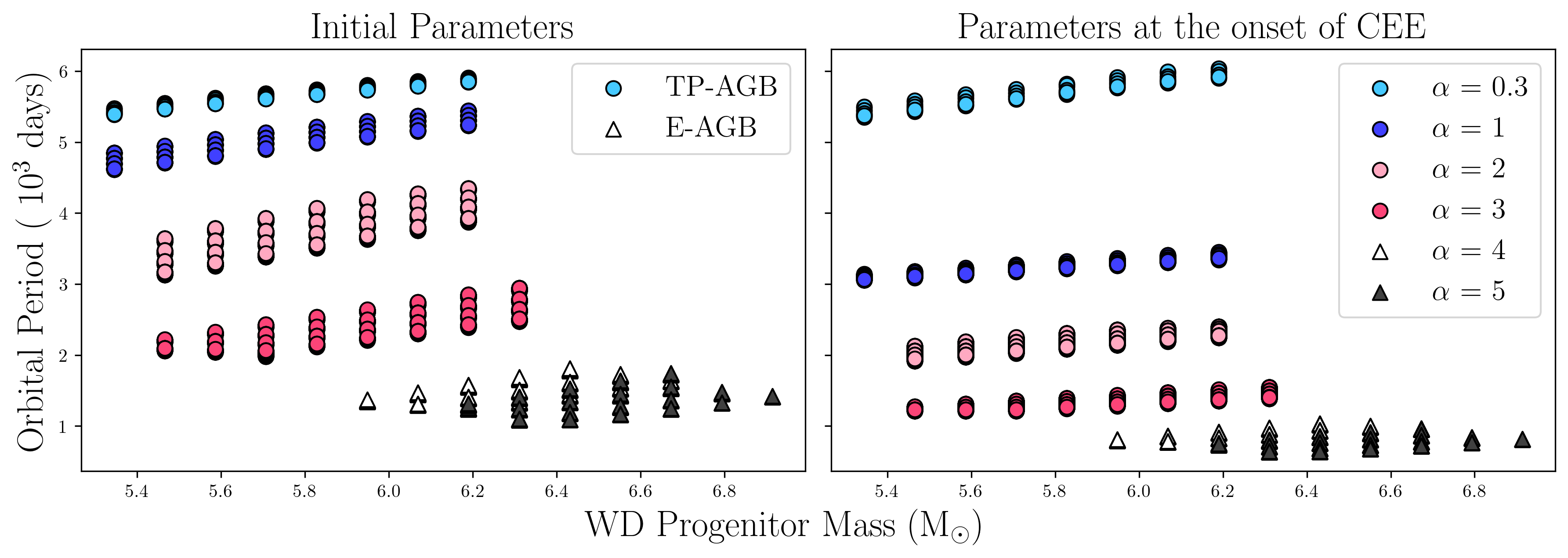}
    \caption{Possible progenitor orbital periods and  WD masses at both $t=0$ (left panel) and the time that CEE occurred (right panel) produced by COSMIC. The colors correspond to the value of $\alpha$ used to evolve the system. The marker shape indicates the stage of evolution of the WD progenitor at the time that CEE occurred, with triangles and circles corresponding to the E-AGB and TP-AGB, respectively.  \label{fig:cosmic2}}
\end{figure*}

\begin{figure}[ht!]
    \includegraphics[width=\linewidth]{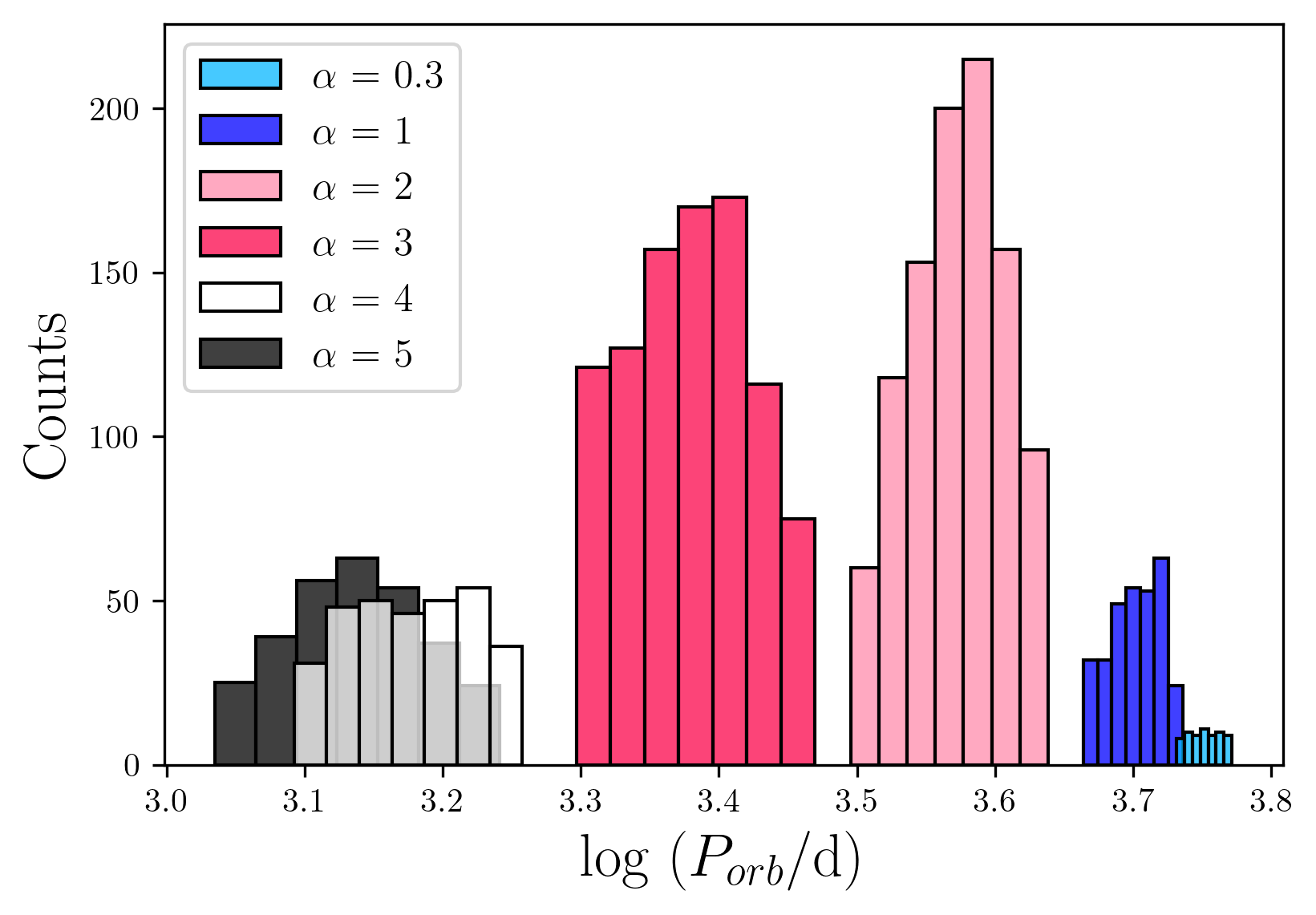}
    \caption{Distribution of initial orbital periods produced by COSMIC for progenitor systems that evolve to produce binaries with properties similar to 2M07515777, colored by the value of $\alpha$ (common envelope efficiency parameter) used in the models. 
    \label{fig:comsic1}}
\end{figure}

Within COSMIC, numerous variables and evolutionary models can be set to best model a system. For our models of 2M07515777, we set \texttt{qcflag} = 5, which uses equations from Section 2.3 of \citet{2019MNRAS.490.3740N} to calculate critical mass ratios for the onset of unstable mass transfer and a common envelope during RLOF. We varied the value of the common envelope ejection efficiency parameter $\alpha$ from 0.3-5. Otherwise, we evolved the above grid of possible progenitor systems for 1.5 Gyr (the estimated cooling age of the system; Section \ref{subsubsec:timescale}). We keep all other variables as their default\footnote{\url{https://github.com/COSMIC-PopSynth/COSMIC/blob/develop/examples/Params.ini}}
 except for \texttt{xi} = 1.0, which assumes that wind lost from the primary transfers angular momentum to the secondary and \texttt{ceflag} = 0, which assumes that only the core mass of the donor star in a common envelope will be used to calculate the orbital energy at common envelope onset. This means that alphas will be slightly higher when compared to \texttt{ceflag} = 1 where the total mass of the donor will be used. We chose the standard value of $\alpha=0.3$ (\citet{2010A&A...520A..86Z}) as the starting point and increased the value in increments to $\alpha =5$. Our common envelope ejection efficiency choices were chosen to explore a wide range of orbital periods at which the common envelope could be initiated. If the common envelope is initiated at a wider orbital period, the orbital energy reservoir is larger and can thus be used with lower ejection efficiencies to unbind the envelope successfully and produce a WD binary with the present day orbital period of the heavy WD. Conversely, higher ejection efficiencies are needed in cases where the heavy WD progenitor initiates a common envelope in an orbit with lower orbital periods and thus smaller orbital energy reservoirs.

We then selected all resulting binary systems with similar characteristics to 2M07515777. The systems we consider in subsequent analyses exist in the following parameter space: P=10.2-10.4 days, $M_{wd}=1.02$-1.15~$M_{\odot}$, $M_{ms}=0.59$-0.75~$M_{\odot}$, $T_{eff, ms}$ = 4000-4600~K, and $R_{ms}=0.59$-0.73~$R_{\odot}$. These ranges reflect the values and corresponding uncertainties for the stellar properties of 2M07515777 shown in Table \ref{tab:all_param}, with the exception of a broader range of periods to account for the assumptions and simplifications made within the applied COMSIC prescriptions (including the assumption of a circular orbit with $e=0$).

\begin{figure*}[ht!]
\centering
\includegraphics[width=.9\linewidth]{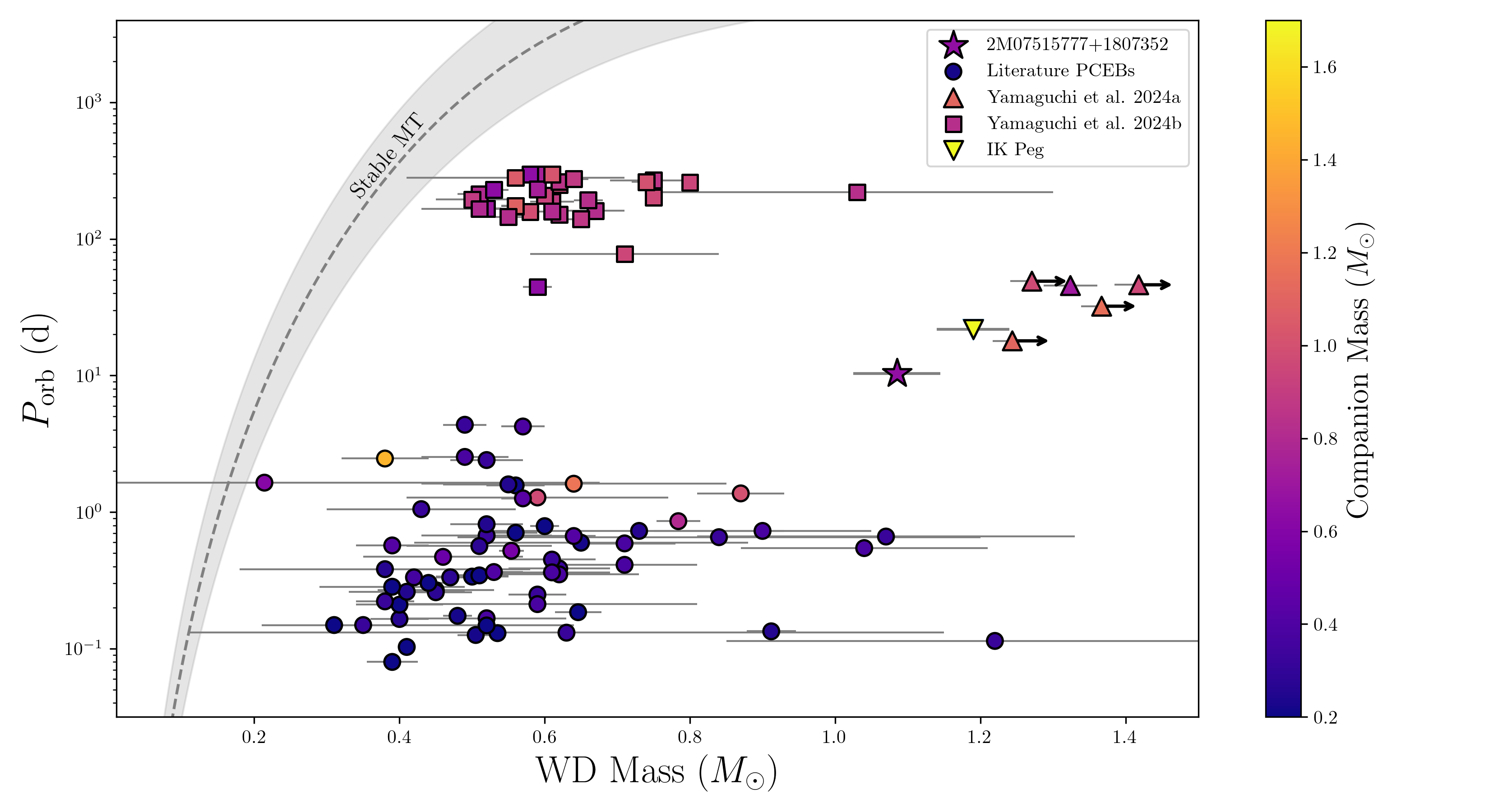}
    \caption{ Orbital period (days) vs. WD mass (M$_\odot$) for PCEBs, with each point colored according to the MS companion mass (adapted from \citet{ 2024MNRAS.52711719Y}). We show 2M07515777 (star marker), along with the other known massive WD systems (triangle markers; the arrows indicate lower limits) including IK Peg (\citet{1993MNRAS.262..277W}) and those presented by \citet{2024PASP..136h4202Y}. Additionally, we plot the sample of wide PCEBs found by \cite{2024MNRAS.52711719Y}. Finally, we plot literature PCEBs (circle markers) including those discovered by \citet{2007MNRAS.382.1377R} using SDSS and later compiled with other known systems by \citet{2010A&A...520A..86Z}, as well as PCEBs found by the WD Binary Pathways Survey \citep{2021MNRAS.501.1677H, 2022MNRAS.512.1843H, 2022MNRAS.517.2867H}. The dashed line illustrates the relation between final orbital period and WD mass expected at the end of stable MT (with a spread in orbital period of a factor of $\sim$$2.4$), taken from \cite{1995MNRAS.273..731R}. 
    \label{fig:diss}}
\end{figure*}

Figure \ref{fig:cosmic2} shows the values of orbital period and progenitor primary mass both at the initial state of the system and at the onset of CEE, as determined by COSMIC. The distribution of potential initial orbital periods ranges from approximately log$(P)=3$ to log$(P)=3.8$ days, which agrees well with the period range that is expected to produce wide ($P_{orb}\sim$ 10-1000 days) PCEBs through CEE while the progenitor primary was on the AGB \citep{2024PASP..136h4202Y}. All possible initial systems produced by COSMIC formed while the WD progenitor was on the AGB, either on the early (E)-AGB or thermally pulsing (TP)-AGB. If 2M07515777 underwent CEE while the WD progenitor was on the TP-AGB, we find that the system likely had an initial orbital period between $\sim$2000-6000 days and a WD progenitor mass between $\sim$5.3-6.4 M$_{\odot}$. Alternatively, if CEE occurred while the WD progenitor was still on the E-AGB, we find that the system likely had an initial orbital period between $\sim$1100-1800 days and a WD progenitor mass between $\sim$5.9-6.8 M$_{\odot}$. We discuss these scenarios further in Section \ref{form}. 

The resulting initial period distribution of possible progenitor systems for 2M07515777 is shown in Figure \ref{fig:comsic1}. Of the uniform grid search that we performed, 2M07515777 was most frequently produced when the progenitor was on the TP-AGB. We also find that for any value of $\alpha$ that we set, COSMIC predicts that CEE occurred $\sim$60-100 Myr into its evolution. At the onset of CEE, we find that the orbital period was reduced to $P\lesssim 3500$ days for $\alpha$ $\gtrsim1$ as seen in Figure \ref{fig:cosmic2}.

\section{Discussion}\label{sec:discussion}

\subsection{PCEB Parameter Space}
2M07515777 hosts a WD that is among the most massive WD companions known to exist in identified PCEBs. Its orbital properties align remarkably well with other PCEBs hosting massive WDs, specifically the systems presented by \citet{2024PASP..136h4202Y}, which not only host massive WDs ($\gtrsim$~1.1 M$_\odot$), but also have long ($>$~10 days) post-CEE orbital periods ranging from 18-49 days and slightly non-zero eccentricities. IK Peg \citep{1993MNRAS.262..277W} is yet another example of a long-period (22 days) PCEB hosting a massive WD ($\sim$1.2 M$_\odot$). 

We show the WD mass and orbital period of 2M07515777 in comparison to these massive WD systems, along with other literature PCEBs, in Figure \ref{fig:diss}. As shown in the figure, 2M07515777 further populates a sparsely occupied region of PCEB parameter space in which lies an isolated group of other known massive WDs in PCEBs, which may suggest a common formation pathway for systems with these parameters. Despite 2M07515777 existing in the same domain as this group, it importantly has both a lower orbital period and WD mass than the others.

Although there are several other systems shown in Figure \ref{fig:diss} that have similar WD masses to 2M07515777 at smaller orbital periods, they all have large mass uncertainties. Additionally, since these systems all have periods of $\lesssim$1 day, their formation pathways likely differ from 2M07515777 in that they can more easily be explained solely by released orbital energy via the general $\alpha$-formalism \citep{2024A&A...687A..12B} in contrast to perhaps needing extra energy from alternative sources.

\subsection{Wide orbit and the efficiency of envelope ejection}

\label{sec:eff}
Traditional PCEBs are commonly thought to exhibit orbital periods ranging from hours to a few days \citep{1993PASP..105.1373I}. Thus, 2M07515777 is in a wide orbit for a PCEB, making this system a valuable case for defining the period distribution produced through CEE as well as the amount of released orbital energy needed for envelope ejection in long-period systems. 

The long period of this system may suggest that additional energy (e.g., recombination energy or accretion energy), beyond the liberated orbital energy from inward spiral, was required for envelope ejection. The role and importance of extra energy sources in producing long-period PCEBs like 2M07515777 has yet to be agreed upon in the literature. For example, IK Peg has previously been found to require extra energy to explain its 22-day period \citep{2010MNRAS.403..179D,2010A&A...520A..86Z}. Additionally, \citet{2024PASP..136h4202Y,2024MNRAS.52711719Y} find that recombination energy can indeed play an important role in the evolution of the common envelope and should not be neglected when modeling PCEB orbits, particularity those in wide separations. 

Some studies propose accretion energy as a mechanism to supply the extra energy needed to eject the common envelope \citep{2025RAA....25b5023S, 2025RAA....25k5014W}. In this scenario, a pair of jets are launched from an accretion disk that forms after the MS companion accretes material. The energy liberated by accretion may then couple to the envelope and assist in envelope ejection. Although there are no direct observations on the existence of jets during common-envelope events, the morphology of some planetary nebulae hosting central post-CEE binary systems have been interpreted as indirect evidence that jets may help shape the ejecta \citep{2020cee..book.....I, 2014MNRAS.440L..16B}. \citet{2014MNRAS.440L..16B} note that jets may allow accretion at a high rate to the MS companion. Considering that the MS star in 2M07515777 appears to be rapidly rotating (Section \ref{subsubsec:rot}), which can be interpreted as an indicator of accretion), this may suggest that accretion associated with jets may have played a role in the evolution of this system.

However, \citet{2024A&A...687A..12B} argue that the orbital properties of all long-period PCEBs containing massive WDs (including systems with periods up to 1000 days, but specifically referring to the six systems in Figure \ref{fig:diss} besides 2M07515777) can be explained without the need for an extra energy source, even for inefficient envelope ejection (i.e. systems with $\alpha$ as low as 0.2). They attribute previous findings of recombination energy being required for long-period systems to simplified assumptions in the computed models. 

The existence of 2M07515777 is predicted by the post-CE binary population synthesis carried out by \citet{2024A&A...687A..12B}. Specifically, their models predict an over-density of systems with orbital periods between $\sim$1-10 days that also host WDs of masses $\gtrsim 0.9$ M$_{\odot}$ (see Figure 1 of \citet{2024A&A...687A..12B}). This accumulation appears in their models for any value of $\alpha$, which could suggest that 2M07515777 may have been produced without recombination energy. However, given the entirety of current literature, it is still uncertain whether or not recombination energy played a role at all during the formation of 2M0751577. 

\citet{2024A&A...687A..12B} also produce models for the maximum possible orbital period as a function of the WD mass for varying zero-age companion masses. The properties of 2M07515777 agree well with these models, as the system (with a MS companion mass of $0.66 \pm 0.08$ M$_{\odot}$) lies well below $\sim1000$ days, which is the predicted maximum orbital period for companion masses $\sim$0.5-1 M$_{\odot}$. 

2M07515777 (along with the other six massive WD systems) has a longer orbital period than the maximum periods predicted by \citet{2010A&A...520A..86Z}, which do not account for WD progenitors on the TP-AGB and therefore result in much shorter maximum periods than \citet{2024A&A...687A..12B} for massive WD systems. \citet{2024A&A...687A..12B} also found that when they use the algorithm from Zorotovic et al. while allowing for TP-AGB progenitors, the resulting maximum orbital periods agree well with their own models. This demonstrates the importance of considering the TP-AGB phase in the formation of wide, post-CE systems hosting massive WDs. We consider formation pathways for 2M07515777 in greater detail in Section \ref{form}.    

\subsection{Non-zero Eccentricity}
 Post-CE orbital eccentricities can be useful diagnostics for better understanding the final phases of CEE. PCEBs are generally expected to have circularized orbits as a result of the tidal interactions that occur prior to CEE and during the dynamical plunge-in of the secondary \citep{2013A&ARv..21...59I}. However, residual eccentricity might be present due to envelope ejection immediately following a dynamical plunge-in (it is expected that more eccentricity would be lost during a slower spiral-in), accretion streams, or gravitational interactions with a nonaxisymmetric eccentric envelope \citep{2013A&ARv..21...59I, 2023A&A...674A.121G}. An example of this residual eccentricity, a $\sim$15 day WD+MS binary with $e$$\simeq$0.068 \citep{1999A&A...344..897D}, is highlighted by \citet{2013A&ARv..21...59I}. 3D hydrodynamic simulations have also found non-zero ($e$ $\lesssim$ 0.1) eccentricities at the end of CEE as a result of initial eccentricity, the density profile of envelope material, or the physics of gaseous dynamical friction \citep{2021MNRAS.507.2659G,2012ApJ...746...74R, 2022MNRAS.513.5465S}. 
These simulation results agree well with the distribution of eccentricities for long-period, WD + AFGK-type systems discussed in \citet{2024PASP..136h4202Y} and \citet{2024A&A...687A..12B} as well as the eccentricities found by \citet{2021ApJ...920...86K}. These findings suggest that the very small eccentricity of 2M07515777  also is a remnant from CEE \citep{2024MNRAS.52711719Y,2013A&ARv..21...59I}. 

\subsection{WD Radius}
Using SED fitting (Section \ref{fig:sed}), we measure a WD radius of $1.54~R_{\earth} \pm 0.07~R_{\earth}$. This WD radius is approximately $12\sigma$ larger than the predicted value using current evolutionary models of ultra-massive WDs, which predict a 1.1~M$_{\odot}$ WD to have a radius of $0.7~R_{\earth}$ for [Fe/H]=0.02 \citep{2005A&A...441..689A}. 

We speculate in the next subsection that the over-sized radius could be explained by an extended helium envelope surrounding the core of the WD, but the structure and stability timescale of the envelope is uncertain and we have thus far been unable to produce a model of such a system.   

\subsection{Formation Pathways}\label{form}
The long orbital period along with the substantial mass of the WD companion indicates that the onset of unstable MT occurred when the progenitor primary was on the AGB. As determined in Section \ref{cosmic}, the progenitor of the massive WD likely had a mass of $\sim$$6$ M$_\odot$, which coupled with minimal mass loss for such stars and a main-sequence companion mass of $\sim$$0.66$ M$_\odot$, gives a mass ratio of approximately 9 when the WD progenitor reached the AGB. This value exceeds the critical mass ratio ($q_{crit}\sim$$3 $-$ 4$; e.g. \citet{1987ApJ...318..794H,2010AIPC.1314...53G,2015ApJ...812...40G,2020ApJ...899..132G,2023A&A...669A..45T}), which sets a conservative lower limit for when unstable AGB MT will occur. Thus the MT forming 2M07515777 was likely to have been dynamically unstable, resulting in the onset of CEE. Here, we consider whether the system was formed through CEE while the WD progenitor was on the E-AGB or TP-AGB. For both proposed pathways, we infer that accretion occurred during the system's evolution, as evidenced by the super-synchronous rotation of the MS companion (Section \ref{subsubsec:rot}. 

\subsubsection{Early-AGB}

The large WD radius and our conjecture of an extended envelope motivates exploration of an E-AGB formation pathway. The COSMIC models discussed in Section \ref{cosmic} allow that 2M07515777 could have formed when the WD progenitor was on the E-AGB with a mass between $\sim$5.8-6.7 M$_{\odot}$ (corresponding to zero-age main sequence WD progenitor masses between $\sim$5.9-6.9 M$_{\odot}$) and orbital period P $\lesssim 1000$ days. In this mass and orbital period range, once the system reached the E-AGB, the binary could have experienced CEE following unstable MT onto the 0.6 M$_{\odot}$ companion. This CEE event would cause orbital shrinkage and the removal of the H-rich envelope, leaving behind a remnant consisting of a C/O core and a helium envelope of several tenths of a solar mass. Then, during the He shell-burning phase that follows, this helium envelope expands and may trigger a second (and potentially stable) episode of RLOF. Following the end of helium shell burning, there will be a C/O core surrounded by a residual helium envelope. Depending on the structure and size of envelope, this may explain the unusually large radius of the WD. 

\subsubsection{Thermally Pulsing-AGB} 

Our COSMIC models found the most frequent formation pathway was through the TP-AGB. Current theory suggests that if a system undergoes CEE while on the AGB, it is most likely to be triggered during the TP-AGB \citep{2024MNRAS.52711719Y,2022MNRAS.517.3181G}. This is due to the radius of the star greatly increasing during thermal pulses and therefore making the envelope less bound, less massive, and easier to eject. 

\citet{2024A&A...687A..12B} state that their predicted over-density of systems with orbital periods between $\sim$1-10 days hosting WDs more massive than $\sim0.9$ M$_{\odot}$ (see Section \ref{sec:eff}) experienced CEE when the WD progenitor was still ascending the AGB or had just reached the TP-AGB. They further suggest that for systems with orbital periods $\gtrsim10$ days (including the other six known wide massive WD systems shown in Figure \ref{sec:discussion}), CEE was triggered when the WD progenitor was already an evolved TP-AGB star with an even less massive and less bound envelope. 

Since 2M07515777 has an orbital period of $\sim 10$ days (and thus lies on the edge of the over-density), the system could have experienced CEE during the early stages of the TP-AGB. At this time, the envelope would likely require more efficient envelope ejection and result in a shorter final orbital period ($\sim10$ days) than systems experiencing CEE when the WD progenitor was a more evolved TP-AGB star (e.g. the other massive WD systems that have periods ranging from $\sim18-49$ days). 

\section{Conclusions}\label{sec:conclusions}
We provide a detailed analysis of 2M07515777\newline+1807352, a likely PCEB identified from excess GALEX FUV and NUV fluxes. From our NEID RV study, we find a 10.3-d orbital period, a slightly non-zero eccentricity, and a massive WD companion ($\gtrsim 1.085$ M$_{\odot}$). We also find, through both SED and \textit{TESS} light-curve analyses, that the massive WD has a radius of $ 1.54~R_{\oplus} \pm 0.07~R_{\oplus}$, $12\sigma$ larger than theoretically expected from WD mass-radius relationships. Additionally, both the Lomb-Scargle analysis and the $v \sin{i}$ of the system indicate the MS companion to be super-syncronously rotating at a period of $\sim$6 days, which may suggest that mass accretion occurred at some point during the evolution of the system.

 Interestingly 2M07515777 shares similar physical characteristics with six other post-CE systems hosting massive WDs, which may point to a common formation pathway. This system also lies within the over-density of systems predicted by \citet{2024A&A...687A..12B} that exhibit orbital periods between $\sim$1-10 days and also host WDs of masses $\gtrsim 0.9$ M$_{\odot}$. We infer that the system likely formed through a phase of CEE, following the onset of MT when the progenitor primary was on the AGB. Specifically, we suggest formation channels in which CEE was triggered either during the E-AGB or early stages of the TP-AGB.  

Though we cannot yet explain the origin of the large WD radius with certainty, we plan to conduct further analysis on the system following the acquisition of both far-UV multi-band photometry and far-UV Lyman-alpha spectroscopy observations with \textit{HST} for this system. These data can confirm the presence of the WD companion and independently measure the WD mass and radius with surface-gravity measurements. 

The analysis of 2M07515777 is part of a larger ongoing study that aims to characterize an extensive sample of PCEBs across the Hertzsprung-Russell diagram. 

\section{Acknowledgments}

This study contains data taken with the NEID instrument, which was funded by the NASA-NSF Exoplanet Observational Research partnership and built by the Pennsylvania State University. NEID is installed on the WIYN telescope, which is operated by the NOIRLab (the National Optical-Infrared Astronomy Research Laboratory), and the NEID archive is operated by the NASA Exoplanet Science Institute at the California Institute of Technology. This work would not have been possible without the incredible staff at WIYN/NEID. 

This work includes data from the European Space
Agency (ESA) mission Gaia (\url{https://www.cosmos.esa.int/
gaia}), processed by the Gaia Data Processing and Analysis
Consortium (DPAC; \url{https://www.cosmos.esa.int/web/gaia/
dpac/} consortium). Funding for the DPAC has been provided
by national institutions, in particular the institutions participating in the Gaia Multilateral Agreement. This paper also utilizes data collected by the TESS mission. Funding for the TESS mission is provided by the NASA’s Science Mission
Directorate.

Some of the data presented in this paper were obtained from the Mikulski Archive for Space Telescopes (MAST) at the Space Telescope Science Institute. The specific observations analyzed can be accessed via \dataset[https://doi.org/10.17909/T9DG6F]{https://doi.org/10.17909/T9DG6F}. STScI is operated by the Association of Universities for Research in Astronomy, Inc., under NASA contract NAS5–26555. Support to MAST for these data is provided by the NASA Office of Space Science via grant NAG5–7584 and by other grants and contracts.

The authors would like to express their gratitude to the following people: Wes Tobin for his expertise and assistance with the \texttt{PHOEBE} software, Andrew Vanderburg and Melinda Soares-Furtado for their initial light-curve analysis of the system, Bill Wolf and Elaina Plonis for their efforts to model the white dwarf envelope, James Davidson and Elliott Horch for their work collecting and reducing the speckle imaging data, Max Kroft for his continual guidance and helpful insight, Andrew Nine for sharing his expertise of stellar abundances, and finally Alison Sills, Emily Leiner, the AGGC team and attendees of the University of Wisconsin-Madison Stars Coffee, and the AGGC team for many valuable discussions regarding the evolutionary pathway and astrophysical implications of the system. 

We acknowledge the support of the National Science Foundation through awards AST-1714506, AST-2307864, AWD-005972, and the Wisconsin Alumni Reserch Fund. We also received support from a Fluno Fellowship.

This work was conducted at the University of Wisconsin-Madison, which is located on occupied ancestral land of the Ho-Chunk people, a place their nation has called Teejop since time immemorial. In an 1832 treaty, the Ho-Chunk were forced
to cede this territory. The university was founded on and funded through this seized land; this legacy enabled the science presented here. Observations for this work were conducted at the WIYN telescope on Kitt Peak, which is part of the lands of the Tohono O’odham Nation.

\textit{Facilities}
NEID, Gaia, TESS, GALEX.

\textit{Software}
\texttt{Astropy} \citep{2013A&A...558A..33A,2018AJ....156..123A,2022ApJ...935..167A}, \texttt{Lightkurve} \citep{2018ascl.soft12013L}, \texttt{pandas} \citep{mckinney-proc-scipy-2010}, \texttt{SciPy} \citep{2020NatMe..17..261V}, \texttt{PHOEBE} \citep{2011ascl.soft06002P}, \texttt{NumPy} \citep{2020Natur.585..357H}, \texttt{Keplersplinev2} (\url{https://github.com/avanderburg/keplersplinev2}), \texttt{emcee} \citep{2013PASP..125..306F,2013ascl.soft03002F, 2019JOSS....4.1864F}, \texttt{ellc} \citep{2016A&A...591A.111M}, \texttt{COSMIC} \citep{2020ApJ...898...71B, 2021ascl.soft08022B}, and VizieR catalog access tool \citep{2000A&AS..143...23O}. 

\clearpage
\bibliography{main}{}

@ARTICLE{2020ApJS..250...34C,
       author = {{Conroy}, Kyle E. and {Kochoska}, Angela and {Hey}, Daniel and {Pablo}, Herbert and {Hambleton}, Kelly M. and {Jones}, David and {Giammarco}, Joseph and {Abdul-Masih}, Michael and {Pr{\v{s}}a}, Andrej},
        title = "{Physics of Eclipsing Binaries. V. General Framework for Solving the Inverse Problem}",
      journal = {\apjs},
         year = 2020,
        month = oct,
       volume = {250},
       number = {2},
          eid = {34},
        pages = {34},
          doi = {10.3847/1538-4365/abb4e2},
archivePrefix = {arXiv},
       eprint = {2006.16951},
 primaryClass = {astro-ph.SR},
       adsurl = {https://ui.adsabs.harvard.edu/abs/2020ApJS..250...34C}
}

@ARTICLE{2021AJ....161..190G,
       author = {{Geller}, Aaron M. and {Mathieu}, Robert D. and {Latham}, David W. and {Pollack}, Maxwell and {Torres}, Guillermo and {Leiner}, Emily M.},
        title = "{Stellar Radial Velocities in the Old Open Cluster M67 (NGC 2682). II. The Spectroscopic Binary Population}",
      journal = {\aj},
         year = 2021,
        month = apr,
       volume = {161},
       number = {4},
          eid = {190},
        pages = {190},
          doi = {10.3847/1538-3881/abdd23},
archivePrefix = {arXiv},
       eprint = {2101.07883},
 primaryClass = {astro-ph.SR},
       adsurl = {https://ui.adsabs.harvard.edu/abs/2021AJ....161..190G}
}

@ARTICLE{2002AJ....124.1144L,
       author = {{Latham}, David W. and {Stefanik}, Robert P. and {Torres}, Guillermo and {Davis}, Robert J. and {Mazeh}, Tsevi and {Carney}, Bruce W. and {Laird}, John B. and {Morse}, Jon A.},
        title = "{A Survey of Proper-Motion Stars. XVI. Orbital Solutions for 171 Single-lined Spectroscopic Binaries}",
      journal = {\aj},
         year = 2002,
        month = aug,
       volume = {124},
       number = {2},
        pages = {1144-1161},
          doi = {10.1086/341384},
       adsurl = {https://ui.adsabs.harvard.edu/abs/2002AJ....124.1144L}
}

@misc{2018yCat.1345....0G,
       author = {{Gaia Collaboration}},
        title = "{VizieR Online Data Catalog: Gaia DR2 (Gaia Collaboration, 2018)}",
 howpublished = {VizieR On-line Data Catalog: I/345.  Originally published in: 2018A\&A...616A...1G; doi:10.5270/esa-ycs},
         year = 2018,
        month = apr,
          eid = {I/345},
          doi = {10.26093/cds/vizier.1345},
       adsurl = {https://ui.adsabs.harvard.edu/abs/2018yCat.1345....0G}
}

@ARTICLE{2022AJ....164..126A,
       author = {{Anguiano}, Borja and {Majewski}, Steven R. and {Stassun}, Keivan G. and {Badenes}, Carles and {Daher}, Christine Mazzola and {Dixon}, Don and {Allende Prieto}, Carlos and {Schneider}, Donald P. and {Price-Whelan}, Adrian M. and {Beaton}, Rachael L.},
        title = "{White Dwarf Binaries across the H-R Diagram}",
      journal = {\aj},
         year = 2022,
        month = oct,
       volume = {164},
       number = {4},
          eid = {126},
        pages = {126},
          doi = {10.3847/1538-3881/ac8357},
archivePrefix = {arXiv},
       eprint = {2207.13992},
 primaryClass = {astro-ph.SR},
       adsurl = {https://ui.adsabs.harvard.edu/abs/2022AJ....164..126A}
}

@ARTICLE{2017ApJS..230...24B,
       author = {{Bianchi}, Luciana and {Shiao}, Bernie and {Thilker}, David},
        title = "{Revised Catalog of GALEX Ultraviolet Sources. I. The All-Sky Survey: GUVcat\_AIS}",
      journal = {\apjs},
         year = 2017,
        month = jun,
       volume = {230},
       number = {2},
          eid = {24},
        pages = {24},
          doi = {10.3847/1538-4365/aa7053},
archivePrefix = {arXiv},
       eprint = {1704.05903},
 primaryClass = {astro-ph.GA},
       adsurl = {https://ui.adsabs.harvard.edu/abs/2017ApJS..230...24B}
}

@ARTICLE{2017AJ....154...94M,
       author = {{Majewski}, Steven R. and {Schiavon}, Ricardo P. and {Frinchaboy}, Peter M. and {Allende Prieto}, Carlos and {Barkhouser}, Robert and {Bizyaev}, Dmitry and {Blank}, Basil and {Brunner}, Sophia and {Burton}, Adam and {Carrera}, Ricardo and {Chojnowski}, S. Drew and {Cunha}, K{\'a}tia and {Epstein}, Courtney and {Fitzgerald}, Greg and {Garc{\'\i}a P{\'e}rez}, Ana E. and {Hearty}, Fred R. and {Henderson}, Chuck and {Holtzman}, Jon A. and {Johnson}, Jennifer A. and {Lam}, Charles R. and {Lawler}, James E. and {Maseman}, Paul and {M{\'e}sz{\'a}ros}, Szabolcs and {Nelson}, Matthew and {Nguyen}, Duy Coung and {Nidever}, David L. and {Pinsonneault}, Marc and {Shetrone}, Matthew and {Smee}, Stephen and {Smith}, Verne V. and {Stolberg}, Todd and {Skrutskie}, Michael F. and {Walker}, Eric and {Wilson}, John C. and {Zasowski}, Gail and {Anders}, Friedrich and {Basu}, Sarbani and {Beland}, Stephane and {Blanton}, Michael R. and {Bovy}, Jo and {Brownstein}, Joel R. and {Carlberg}, Joleen and {Chaplin}, William and {Chiappini}, Cristina and {Eisenstein}, Daniel J. and {Elsworth}, Yvonne and {Feuillet}, Diane and {Fleming}, Scott W. and {Galbraith-Frew}, Jessica and {Garc{\'\i}a}, Rafael A. and {Garc{\'\i}a-Hern{\'a}ndez}, D. An{\'\i}bal and {Gillespie}, Bruce A. and {Girardi}, L{\'e}o and {Gunn}, James E. and {Hasselquist}, Sten and {Hayden}, Michael R. and {Hekker}, Saskia and {Ivans}, Inese and {Kinemuchi}, Karen and {Klaene}, Mark and {Mahadevan}, Suvrath and {Mathur}, Savita and {Mosser}, Beno{\^\i}t and {Muna}, Demitri and {Munn}, Jeffrey A. and {Nichol}, Robert C. and {O'Connell}, Robert W. and {Parejko}, John K. and {Robin}, A.~C. and {Rocha-Pinto}, Helio and {Schultheis}, Matthias and {Serenelli}, Aldo M. and {Shane}, Neville and {Silva Aguirre}, Victor and {Sobeck}, Jennifer S. and {Thompson}, Benjamin and {Troup}, Nicholas W. and {Weinberg}, David H. and {Zamora}, Olga},
        title = "{The Apache Point Observatory Galactic Evolution Experiment (APOGEE)}",
      journal = {\aj},
         year = 2017,
        month = sep,
       volume = {154},
       number = {3},
          eid = {94},
        pages = {94},
          doi = {10.3847/1538-3881/aa784d},
archivePrefix = {arXiv},
       eprint = {1509.05420},
 primaryClass = {astro-ph.IM},
       adsurl = {https://ui.adsabs.harvard.edu/abs/2017AJ....154...94M}
}

@ARTICLE{2018ApJ...866...21C,
       author = {{Cummings}, Jeffrey D. and {Kalirai}, Jason S. and {Tremblay}, P. -E. and {Ramirez-Ruiz}, Enrico and {Choi}, Jieun},
        title = "{The White Dwarf Initial-Final Mass Relation for Progenitor Stars from 0.85 to 7.5 M $_{{\ensuremath{\odot}}}$}",
      journal = {\apj},
         year = 2018,
        month = oct,
       volume = {866},
       number = {1},
          eid = {21},
        pages = {21},
          doi = {10.3847/1538-4357/aadfd6},
archivePrefix = {arXiv},
       eprint = {1809.01673},
 primaryClass = {astro-ph.SR},
       adsurl = {https://ui.adsabs.harvard.edu/abs/2018ApJ...866...21C}
}

@ARTICLE{2024A&A...687A..12B,
       author = {{Belloni}, Diogo and {Schreiber}, Matthias R. and {Zorotovic}, Monica},
        title = "{Formation of long-period post-common-envelope binaries. II. Explaining the self-lensing binary KOI 3278}",
      journal = {\aap},
         year = 2024,
        month = jul,
       volume = {687},
          eid = {A12},
        pages = {A12},
          doi = {10.1051/0004-6361/202449320},
archivePrefix = {arXiv},
       eprint = {2401.17510},
 primaryClass = {astro-ph.SR},
       adsurl = {https://ui.adsabs.harvard.edu/abs/2024A&A...687A..12B}
}

@ARTICLE{StassunTorres:2021,
       author = {{Stassun}, Keivan G. and {Torres}, Guillermo},
        title = "{Parallax Systematics and Photocenter Motions of Benchmark Eclipsing Binaries in Gaia EDR3}",
      journal = {\apjl},
         year = 2021,
        month = feb,
       volume = {907},
       number = {2},
          eid = {L33},
        pages = {L33},
          doi = {10.3847/2041-8213/abdaad},
archivePrefix = {arXiv},
       eprint = {2101.03425},
 primaryClass = {astro-ph.SR},
       adsurl = {https://ui.adsabs.harvard.edu/abs/2021ApJ...907L..33S}
}

@ARTICLE{Stassun:2018,
       author = {{Stassun}, Keivan G. and {Corsaro}, Enrico and {Pepper}, Joshua A. and
         {Gaudi}, B. Scott},
        title = "{Empirical Accurate Masses and Radii of Single Stars with TESS and Gaia}",
      journal = {\aj},
         year = 2018,
        month = jan,
       volume = {155},
       number = {1},
          eid = {22},
        pages = {22},
          doi = {10.3847/1538-3881/aa998a},
archivePrefix = {arXiv},
       eprint = {1710.01460},
 primaryClass = {astro-ph.SR},
       adsurl = {https://ui.adsabs.harvard.edu/abs/2018AJ....155...22S}
}

@ARTICLE{Stassun:2017,
       author = {{Stassun}, Keivan G. and {Collins}, Karen A. and {Gaudi}, B. Scott},
        title = "{Accurate Empirical Radii and Masses of Planets and Their Host Stars with Gaia Parallaxes}",
      journal = {\aj},
         year = 2017,
        month = mar,
       volume = {153},
       number = {3},
          eid = {136},
        pages = {136},
          doi = {10.3847/1538-3881/aa5df3},
archivePrefix = {arXiv},
       eprint = {1609.04389},
 primaryClass = {astro-ph.EP},
       adsurl = {https://ui.adsabs.harvard.edu/abs/2017AJ....153..136S}
}

@ARTICLE{Stassun:2016,
       author = {{Stassun}, Keivan G. and {Torres}, Guillermo},
        title = "{Eclipsing Binaries as Benchmarks for Trigonometric Parallaxes in the Gaia Era}",
      journal = {\aj},
         year = 2016,
        month = dec,
       volume = {152},
       number = {6},
          eid = {180},
        pages = {180},
          doi = {10.3847/0004-6256/152/6/180},
archivePrefix = {arXiv},
       eprint = {1609.02579},
 primaryClass = {astro-ph.SR},
       adsurl = {https://ui.adsabs.harvard.edu/abs/2016AJ....152..180S}
}

@ARTICLE{Schlegel:1998,
       author = {{Schlegel}, David J. and {Finkbeiner}, Douglas P. and {Davis}, Marc},
        title = "{Maps of Dust Infrared Emission for Use in Estimation of Reddening and Cosmic Microwave Background Radiation Foregrounds}",
      journal = {\apj},
         year = 1998,
        month = jun,
       volume = {500},
       number = {2},
        pages = {525-553},
          doi = {10.1086/305772},
archivePrefix = {arXiv},
       eprint = {astro-ph/9710327},
 primaryClass = {astro-ph},
       adsurl = {https://ui.adsabs.harvard.edu/abs/1998ApJ...500..525S}
}

@ARTICLE{Husser:2013,
       author = {{Husser}, T. -O. and {Wende-von Berg}, S. and {Dreizler}, S. and {Homeier}, D. and {Reiners}, A. and {Barman}, T. and {Hauschildt}, P.~H.},
        title = "{A new extensive library of PHOENIX stellar atmospheres and synthetic spectra}",
      journal = {\aap},
         year = 2013,
        month = may,
       volume = {553},
          eid = {A6},
        pages = {A6},
          doi = {10.1051/0004-6361/201219058},
archivePrefix = {arXiv},
       eprint = {1303.5632},
 primaryClass = {astro-ph.SR},
       adsurl = {https://ui.adsabs.harvard.edu/abs/2013A&A...553A...6H}
}

@misc{2019yCat..36270004R,
       author = {{Roeser}, S. and {Schilbach}, E.},
        title = "{VizieR Online Data Catalog: Praesepe (NGC 2632) and its tidal tails (Roeser+, 2019)}",
 howpublished = {VizieR On-line Data Catalog: J/A+A/627/A4. Originally published in: 2019A\&A...627A...4R},
         year = 2019,
        month = jun,
          eid = {J/A+A/627/A4},
          doi = {10.26093/cds/vizier.36270004},
       adsurl = {https://ui.adsabs.harvard.edu/abs/2019yCat..36270004R}
}

@ARTICLE{2013ApJ...775...58B,
       author = {{Boesgaard}, Ann Merchant and {Roper}, Brian W. and {Lum}, Michael G.},
        title = "{The Chemical Composition of Praesepe (M44)}",
      journal = {\apj},
         year = 2013,
        month = sep,
       volume = {775},
       number = {1},
          eid = {58},
        pages = {58},
          doi = {10.1088/0004-637X/775/1/58},
archivePrefix = {arXiv},
       eprint = {1308.3476},
 primaryClass = {astro-ph.SR},
       adsurl = {https://ui.adsabs.harvard.edu/abs/2013ApJ...775...58B}
}

@ARTICLE{2013A&ARv..21...59I,
       author = {{Ivanova}, N. and {Justham}, S. and {Chen}, X. and {De Marco}, O. and {Fryer}, C.~L. and {Gaburov}, E. and {Ge}, H. and {Glebbeek}, E. and {Han}, Z. and {Li}, X. -D. and {Lu}, G. and {Marsh}, T. and {Podsiadlowski}, P. and {Potter}, A. and {Soker}, N. and {Taam}, R. and {Tauris}, T.~M. and {van den Heuvel}, E.~P.~J. and {Webbink}, R.~F.},
        title = "{Common envelope evolution: where we stand and how we can move forward}",
      journal = {\aapr},
         year = 2013,
        month = feb,
       volume = {21},
          eid = {59},
        pages = {59},
          doi = {10.1007/s00159-013-0059-2},
archivePrefix = {arXiv},
       eprint = {1209.4302},
 primaryClass = {astro-ph.HE},
       adsurl = {https://ui.adsabs.harvard.edu/abs/2013A&ARv..21...59I}
}

@INPROCEEDINGS{1976IAUS...73...75P,
       author = {{Paczynski}, B.},
        title = "{Common Envelope Binaries}",
    booktitle = {Structure and Evolution of Close Binary Systems},
         year = 1976,
       editor = {{Eggleton}, Peter and {Mitton}, Simon and {Whelan}, John},
       series = {IAU Symposium},
       volume = {73},
        month = jan,
        pages = {75},
       adsurl = {https://ui.adsabs.harvard.edu/abs/1976IAUS...73...75P}
}

@ARTICLE{2010MNRAS.403..179D,
       author = {{Davis}, P.~J. and {Kolb}, U. and {Willems}, B.},
        title = "{A comprehensive population synthesis study of post-common envelope binaries}",
      journal = {\mnras},
         year = 2010,
        month = mar,
       volume = {403},
       number = {1},
        pages = {179-195},
          doi = {10.1111/j.1365-2966.2009.16138.x},
archivePrefix = {arXiv},
       eprint = {0903.4152},
 primaryClass = {astro-ph.SR},
       adsurl = {https://ui.adsabs.harvard.edu/abs/2010MNRAS.403..179D}
}

@ARTICLE{2024arXiv241104563P,
       author = {{Pellouin}, Cl{\'e}ment and {Dvorkin}, Irina and {Lehoucq}, L{\'e}onard},
        title = "{Evolutionary tracks of binary neutron star progenitors across cosmic times}",
      journal = {arXiv e-prints},
         year = 2024,
        month = nov,
          eid = {arXiv:2411.04563},
        pages = {arXiv:2411.04563},
          doi = {10.48550/arXiv.2411.04563},
archivePrefix = {arXiv},
       eprint = {2411.04563},
 primaryClass = {astro-ph.HE},
       adsurl = {https://ui.adsabs.harvard.edu/abs/2024arXiv241104563P}
}

@ARTICLE{2019MNRAS.490.5888L,
       author = {{Lamberts}, Astrid and {Blunt}, Sarah and {Littenberg}, Tyson B. and {Garrison-Kimmel}, Shea and {Kupfer}, Thomas and {Sanderson}, Robyn E.},
        title = "{Predicting the LISA white dwarf binary population in the Milky Way with cosmological simulations}",
      journal = {\mnras},
         year = 2019,
        month = dec,
       volume = {490},
       number = {4},
        pages = {5888-5903},
          doi = {10.1093/mnras/stz2834},
archivePrefix = {arXiv},
       eprint = {1907.00014},
 primaryClass = {astro-ph.HE},
       adsurl = {https://ui.adsabs.harvard.edu/abs/2019MNRAS.490.5888L}
}

@ARTICLE{2016ApJ...830L..18B,
       author = {{Breivik}, Katelyn and {Rodriguez}, Carl L. and {Larson}, Shane L. and {Kalogera}, Vassiliki and {Rasio}, Frederic A.},
        title = "{Distinguishing between Formation Channels for Binary Black Holes with LISA}",
      journal = {\apjl},
         year = 2016,
        month = oct,
       volume = {830},
       number = {1},
          eid = {L18},
        pages = {L18},
          doi = {10.3847/2041-8205/830/1/L18},
archivePrefix = {arXiv},
       eprint = {1606.09558},
 primaryClass = {astro-ph.GA},
       adsurl = {https://ui.adsabs.harvard.edu/abs/2016ApJ...830L..18B}
}

@ARTICLE{1971AJ.....76..544L,
       author = {{Lucy}, L.~B. and {Sweeney}, M.~A.},
        title = "{Spectroscopic binaries with circular orbits.}",
      journal = {\aj},
         year = 1971,
        month = aug,
       volume = {76},
        pages = {544-556},
          doi = {10.1086/111159},
       adsurl = {https://ui.adsabs.harvard.edu/abs/1971AJ.....76..544L}
}

@ARTICLE{1993PASP..105..841L,
       author = {{Landsman}, Wayne and {Simon}, Theodore and {Bergeron}, P.},
        title = "{The Hot White Dwarf Companions of HR 1608, HR 8210, and HD 15638}",
      journal = {\pasp},
         year = 1993,
        month = aug,
       volume = {105},
        pages = {841},
          doi = {10.1086/133242},
       adsurl = {https://ui.adsabs.harvard.edu/abs/1993PASP..105..841L}
}

@ARTICLE{2024A&A...686A..61B,
       author = {{Belloni}, Diogo and {Zorotovic}, Monica and {Schreiber}, Matthias R. and {Parsons}, Steven G. and {Moe}, Maxwell and {Garbutt}, James A.},
        title = "{Formation of long-period post-common envelope binaries. I. No extra energy is needed to explain oxygen-neon white dwarfs paired with AFGK-type main-sequence stars}",
      journal = {\aap},
         year = 2024,
        month = jun,
       volume = {686},
          eid = {A61},
        pages = {A61},
          doi = {10.1051/0004-6361/202449235},
archivePrefix = {arXiv},
       eprint = {2401.07692},
 primaryClass = {astro-ph.SR},
       adsurl = {https://ui.adsabs.harvard.edu/abs/2024A&A...686A..61B}
}

@ARTICLE{2024MNRAS.52711719Y,
       author = {{Yamaguchi}, Natsuko and {El-Badry}, Kareem and {Fuller}, Jim and {Latham}, David W. and {Cargile}, Phillip A. and {Mazeh}, Tsevi and {Shahaf}, Sahar and {Bieryla}, Allyson and {Buchhave}, Lars A. and {Hobson}, Melissa},
        title = "{Wide post-common envelope binaries containing ultramassive white dwarfs: evidence for efficient envelope ejection in massive asymptotic giant branch stars}",
      journal = {\mnras},
         year = 2024,
        month = feb,
       volume = {527},
       number = {4},
        pages = {11719-11739},
          doi = {10.1093/mnras/stad4005},
archivePrefix = {arXiv},
       eprint = {2309.15905},
 primaryClass = {astro-ph.SR},
       adsurl = {https://ui.adsabs.harvard.edu/abs/2024MNRAS.52711719Y}
}

@ARTICLE{1984ApJ...277..355W,
       author = {{Webbink}, R.~F.},
        title = "{Double white dwarfs as progenitors of R Coronae Borealis stars and type I supernovae.}",
      journal = {\apj},
         year = 1984,
        month = feb,
       volume = {277},
        pages = {355-360},
          doi = {10.1086/161701},
       adsurl = {https://ui.adsabs.harvard.edu/abs/1984ApJ...277..355W}
}

@INPROCEEDINGS{2022BAAS...54e4602R,
       author = {{Ricker}, George R. and {Winn}, Josh and {Vanderspek}, Roland},
        title = "{The Transiting Exoplanet Survey Satellite (TESS)}",
    booktitle = {Bulletin of the American Astronomical Society},
         year = 2022,
       volume = {54},
        month = jun,
          eid = {406.02},
        pages = {406.02},
       adsurl = {https://ui.adsabs.harvard.edu/abs/2022BAAS...54e4602R}
}

@ARTICLE{1993PASP..105.1373I,
       author = {{Iben}, Jr., Icko and {Livio}, Mario},
        title = "{Common Envelopes in Binary Star Evolution}",
      journal = {\pasp},
         year = 1993,
        month = dec,
       volume = {105},
        pages = {1373},
          doi = {10.1086/133321},
       adsurl = {https://ui.adsabs.harvard.edu/abs/1993PASP..105.1373I}
}

@ARTICLE{1999A&A...344..897D,
       author = {{Delfosse}, X. and {Forveille}, T. and {Beuzit}, J. -L. and {Udry}, S. and {Mayor}, M. and {Perrier}, C.},
        title = "{New neighbours. I. 13 new companions to nearby M dwarfs}",
      journal = {\aap},
         year = 1999,
        month = apr,
       volume = {344},
        pages = {897-910},
          doi = {10.48550/arXiv.astro-ph/9812008},
archivePrefix = {arXiv},
       eprint = {astro-ph/9812008},
 primaryClass = {astro-ph},
       adsurl = {https://ui.adsabs.harvard.edu/abs/1999A&A...344..897D}
}

@ARTICLE{2023A&A...674A.121G,
       author = {{Gagnier}, Damien and {Pejcha}, Ond{\v{r}}ej},
        title = "{Post-dynamical inspiral phase of common envelope evolution. Binary orbit evolution and angular momentum transport}",
      journal = {\aap},
         year = 2023,
        month = jun,
       volume = {674},
          eid = {A121},
        pages = {A121},
          doi = {10.1051/0004-6361/202346057},
archivePrefix = {arXiv},
       eprint = {2302.00691},
 primaryClass = {astro-ph.SR},
       adsurl = {https://ui.adsabs.harvard.edu/abs/2023A&A...674A.121G}
}

@ARTICLE{2022MNRAS.513.5465S,
       author = {{Sz{\"o}lgy{\'e}n}, {\'A}kos and {MacLeod}, Morgan and {Loeb}, Abraham},
        title = "{Eccentricity evolution in gaseous dynamical friction}",
      journal = {\mnras},
         year = 2022,
        month = jul,
       volume = {513},
       number = {4},
        pages = {5465-5473},
          doi = {10.1093/mnras/stac1294},
archivePrefix = {arXiv},
       eprint = {2203.01334},
 primaryClass = {astro-ph.EP},
       adsurl = {https://ui.adsabs.harvard.edu/abs/2022MNRAS.513.5465S}
}

@ARTICLE{2021MNRAS.507.2659G,
       author = {{Glanz}, Hila and {Perets}, Hagai B.},
        title = "{Common envelope evolution of eccentric binaries}",
      journal = {\mnras},
         year = 2021,
        month = oct,
       volume = {507},
       number = {2},
        pages = {2659-2670},
          doi = {10.1093/mnras/stab2291},
archivePrefix = {arXiv},
       eprint = {2105.02227},
 primaryClass = {astro-ph.SR},
       adsurl = {https://ui.adsabs.harvard.edu/abs/2021MNRAS.507.2659G}
}

@ARTICLE{2012ApJ...746...74R,
       author = {{Ricker}, Paul M. and {Taam}, Ronald E.},
        title = "{An AMR Study of the Common-envelope Phase of Binary Evolution}",
      journal = {\apj},
         year = 2012,
        month = feb,
       volume = {746},
       number = {1},
          eid = {74},
        pages = {74},
          doi = {10.1088/0004-637X/746/1/74},
archivePrefix = {arXiv},
       eprint = {1107.3889},
 primaryClass = {astro-ph.SR},
       adsurl = {https://ui.adsabs.harvard.edu/abs/2012ApJ...746...74R}
}

@ARTICLE{2021ApJ...920...86K,
       author = {{Kruckow}, Matthias U. and {Neunteufel}, Patrick G. and {Di Stefano}, Rosanne and {Gao}, Yan and {Kobayashi}, Chiaki},
        title = "{A Catalog of Potential Post-Common Envelope Binaries}",
      journal = {\apj},
         year = 2021,
        month = oct,
       volume = {920},
       number = {2},
          eid = {86},
        pages = {86},
          doi = {10.3847/1538-4357/ac13ac},
archivePrefix = {arXiv},
       eprint = {2107.05221},
 primaryClass = {astro-ph.SR},
       adsurl = {https://ui.adsabs.harvard.edu/abs/2021ApJ...920...86K}
}

@ARTICLE{2015MNRAS.447.2181I,
       author = {{Ivanova}, N. and {Justham}, S. and {Podsiadlowski}, Ph.},
        title = "{On the role of recombination in common-envelope ejections}",
      journal = {\mnras},
         year = 2015,
        month = mar,
       volume = {447},
       number = {3},
        pages = {2181-2197},
          doi = {10.1093/mnras/stu2582},
archivePrefix = {arXiv},
       eprint = {1409.3260},
 primaryClass = {astro-ph.SR},
       adsurl = {https://ui.adsabs.harvard.edu/abs/2015MNRAS.447.2181I}
}

@ARTICLE{2022MNRAS.516.4669L,
       author = {{Lau}, Mike Y.~M. and {Hirai}, Ryosuke and {Price}, Daniel J. and {Mandel}, Ilya},
        title = "{Common envelopes in massive stars II: The distinct roles of hydrogen and helium recombination}",
      journal = {\mnras},
         year = 2022,
        month = nov,
       volume = {516},
       number = {4},
        pages = {4669-4678},
          doi = {10.1093/mnras/stac2490},
archivePrefix = {arXiv},
       eprint = {2206.06411},
 primaryClass = {astro-ph.SR},
       adsurl = {https://ui.adsabs.harvard.edu/abs/2022MNRAS.516.4669L}
}

@ARTICLE{2012MNRAS.423..320R,
       author = {{Rebassa-Mansergas}, A. and {Zorotovic}, M. and {Schreiber}, M.~R. and {G{\"a}nsicke}, B.~T. and {Southworth}, J. and {Nebot G{\'o}mez-Mor{\'a}n}, A. and {Tappert}, C. and {Koester}, D. and {Pyrzas}, S. and {Papadaki}, C. and {Schmidtobreick}, L. and {Schwope}, A. and {Toloza}, O.},
        title = "{Post-common envelope binaries from SDSS - XVI. Long orbital period systems and the energy budget of common envelope evolution}",
      journal = {\mnras},
         year = 2012,
        month = jun,
       volume = {423},
       number = {1},
        pages = {320-327},
          doi = {10.1111/j.1365-2966.2012.20880.x},
archivePrefix = {arXiv},
       eprint = {1203.1208},
 primaryClass = {astro-ph.SR},
       adsurl = {https://ui.adsabs.harvard.edu/abs/2012MNRAS.423..320R}
}

@INPROCEEDINGS{2008ASSL..352..233W,
       author = {{Webbink}, Ronald F.},
        title = "{Common Envelope Evolution Redux}",
    booktitle = {Astrophysics and Space Science Library},
         year = 2008,
       editor = {{Milone}, Eugene F. and {Leahy}, Denis A. and {Hobill}, David W.},
       series = {Astrophysics and Space Science Library},
       volume = {352},
        month = jan,
        pages = {233},
          doi = {10.1007/978-1-4020-6544-6_13},
archivePrefix = {arXiv},
       eprint = {0704.0280},
 primaryClass = {astro-ph},
       adsurl = {https://ui.adsabs.harvard.edu/abs/2008ASSL..352..233W}
}

@ARTICLE{2019MNRAS.490.2550I,
       author = {{Iaconi}, Roberto and {De Marco}, Orsola},
        title = "{Speaking with one voice: simulations and observations discuss the common envelope {\ensuremath{\alpha}} parameter}",
      journal = {\mnras},
         year = 2019,
        month = dec,
       volume = {490},
       number = {2},
        pages = {2550-2566},
          doi = {10.1093/mnras/stz2756},
archivePrefix = {arXiv},
       eprint = {1902.02039},
 primaryClass = {astro-ph.SR},
       adsurl = {https://ui.adsabs.harvard.edu/abs/2019MNRAS.490.2550I}
}

@ARTICLE{2012ApJ...744...52P,
       author = {{Passy}, Jean-Claude and {De Marco}, Orsola and {Fryer}, Chris L. and {Herwig}, Falk and {Diehl}, Steven and {Oishi}, Jeffrey S. and {Mac Low}, Mordecai-Mark and {Bryan}, Greg L. and {Rockefeller}, Gabriel},
        title = "{Simulating the Common Envelope Phase of a Red Giant Using Smoothed-particle Hydrodynamics and Uniform-grid Codes}",
      journal = {\apj},
         year = 2012,
        month = jan,
       volume = {744},
       number = {1},
          eid = {52},
        pages = {52},
          doi = {10.1088/0004-637X/744/1/52},
archivePrefix = {arXiv},
       eprint = {1107.5072},
 primaryClass = {astro-ph.SR},
       adsurl = {https://ui.adsabs.harvard.edu/abs/2012ApJ...744...52P}
}

@ARTICLE{2016MNRAS.460.3992N,
       author = {{Nandez}, J.~L.~A. and {Ivanova}, N.},
        title = "{Common envelope events with low-mass giants: understanding the energy budget}",
      journal = {\mnras},
         year = 2016,
        month = aug,
       volume = {460},
       number = {4},
        pages = {3992-4002},
          doi = {10.1093/mnras/stw1266},
archivePrefix = {arXiv},
       eprint = {1606.04922},
 primaryClass = {astro-ph.SR},
       adsurl = {https://ui.adsabs.harvard.edu/abs/2016MNRAS.460.3992N}
}

@ARTICLE{2015ApJ...806..135H,
       author = {{Hwang}, J. and {Lombardi}, Jr., J.~C. and {Rasio}, F.~A. and {Kalogera}, V.},
        title = "{Stability and Coalescence of Massive Twin Binaries}",
      journal = {\apj},
         year = 2015,
        month = jun,
       volume = {806},
       number = {1},
          eid = {135},
        pages = {135},
          doi = {10.1088/0004-637X/806/1/135},
archivePrefix = {arXiv},
       eprint = {1505.05812},
 primaryClass = {astro-ph.SR},
       adsurl = {https://ui.adsabs.harvard.edu/abs/2015ApJ...806..135H}
}

@ARTICLE{2020RAA....20..135W,
       author = {{Wang}, Bo and {Liu}, Dongdong},
        title = "{The formation of neutron star systems through accretion-induced collapse in white-dwarf binaries}",
      journal = {Research in Astronomy and Astrophysics},
         year = 2020,
        month = sep,
       volume = {20},
       number = {9},
          eid = {135},
        pages = {135},
          doi = {10.1088/1674-4527/20/9/135},
archivePrefix = {arXiv},
       eprint = {2005.01880},
 primaryClass = {astro-ph.SR},
       adsurl = {https://ui.adsabs.harvard.edu/abs/2020RAA....20..135W}
}

@ARTICLE{1991ApJ...370..709H,
       author = {{Hjellming}, Michael S. and {Taam}, Ronald E.},
        title = "{The Response of Main-Sequence Stars within a Common Envelope}",
      journal = {\apj},
         year = 1991,
        month = apr,
       volume = {370},
        pages = {709},
          doi = {10.1086/169854},
       adsurl = {https://ui.adsabs.harvard.edu/abs/1991ApJ...370..709H}
}

@ARTICLE{2022A&A...663A..39B,
       author = {{Bhat}, A. and {Irrgang}, A. and {Heber}, U.},
        title = "{The origin of early-type runaway stars from open clusters}",
      journal = {\aap},
         year = 2022,
        month = jul,
       volume = {663},
          eid = {A39},
        pages = {A39},
          doi = {10.1051/0004-6361/202142993},
archivePrefix = {arXiv},
       eprint = {2204.01594},
 primaryClass = {astro-ph.SR},
       adsurl = {https://ui.adsabs.harvard.edu/abs/2022A&A...663A..39B}
}

@ARTICLE{2007AJ....134.2340K,
       author = {{Kraus}, Adam L. and {Hillenbrand}, Lynne A.},
        title = "{The Stellar Populations of Praesepe and Coma Berenices}",
      journal = {\aj},
         year = 2007,
        month = dec,
       volume = {134},
       number = {6},
        pages = {2340-2352},
          doi = {10.1086/522831},
archivePrefix = {arXiv},
       eprint = {0708.2719},
 primaryClass = {astro-ph},
       adsurl = {https://ui.adsabs.harvard.edu/abs/2007AJ....134.2340K}
}

@ARTICLE{2019A&A...627A...4R,
       author = {{R{\"o}ser}, Siegfried and {Schilbach}, Elena},
        title = "{Praesepe (NGC 2632) and its tidal tails}",
      journal = {\aap},
         year = 2019,
        month = jul,
       volume = {627},
          eid = {A4},
        pages = {A4},
          doi = {10.1051/0004-6361/201935502},
archivePrefix = {arXiv},
       eprint = {1903.08610},
 primaryClass = {astro-ph.SR},
       adsurl = {https://ui.adsabs.harvard.edu/abs/2019A&A...627A...4R}
}

@ARTICLE{2008MNRAS.390.1635R,
       author = {{Rebassa-Mansergas}, A. and {G{\"a}nsicke}, B.~T. and {Schreiber}, M.~R. and {Southworth}, J. and {Schwope}, A.~D. and {Nebot Gomez-Moran}, A. and {Aungwerojwit}, A. and {Rodr{\'\i}guez-Gil}, P. and {Karamanavis}, V. and {Krumpe}, M. and {Tremou}, E. and {Schwarz}, R. and {Staude}, A. and {Vogel}, J.},
        title = "{Post-common envelope binaries from SDSS - III. Seven new orbital periods}",
      journal = {\mnras},
         year = 2008,
        month = nov,
       volume = {390},
       number = {4},
        pages = {1635-1646},
          doi = {10.1111/j.1365-2966.2008.13850.x},
archivePrefix = {arXiv},
       eprint = {0808.2148},
 primaryClass = {astro-ph},
       adsurl = {https://ui.adsabs.harvard.edu/abs/2008MNRAS.390.1635R}
}

@ARTICLE{2011A&A...536A..43N,
       author = {{Nebot G{\'o}mez-Mor{\'a}n}, A. and {G{\"a}nsicke}, B.~T. and {Schreiber}, M.~R. and {Rebassa-Mansergas}, A. and {Schwope}, A.~D. and {Southworth}, J. and {Aungwerojwit}, A. and {Bothe}, M. and {Davis}, P.~J. and {Kolb}, U. and {M{\"u}ller}, M. and {Papadaki}, C. and {Pyrzas}, S. and {Rabitz}, A. and {Rodr{\'\i}guez-Gil}, P. and {Schmidtobreick}, L. and {Schwarz}, R. and {Tappert}, C. and {Toloza}, O. and {Vogel}, J. and {Zorotovic}, M.},
        title = "{Post common envelope binaries from SDSS. XII. The orbital period distribution}",
      journal = {\aap},
         year = 2011,
        month = dec,
       volume = {536},
          eid = {A43},
        pages = {A43},
          doi = {10.1051/0004-6361/201117514},
archivePrefix = {arXiv},
       eprint = {1109.6662},
 primaryClass = {astro-ph.SR},
       adsurl = {https://ui.adsabs.harvard.edu/abs/2011A&A...536A..43N}
}

@ARTICLE{2019A&A...628A..66L,
       author = {{Lodieu}, N. and {P{\'e}rez-Garrido}, A. and {Smart}, R.~L. and {Silvotti}, R.},
        title = "{A 5D view of the {\ensuremath{\alpha}} Per, Pleiades, and Praesepe clusters}",
      journal = {\aap},
         year = 2019,
        month = aug,
       volume = {628},
          eid = {A66},
        pages = {A66},
          doi = {10.1051/0004-6361/201935533},
archivePrefix = {arXiv},
       eprint = {1906.03924},
 primaryClass = {astro-ph.SR},
       adsurl = {https://ui.adsabs.harvard.edu/abs/2019A&A...628A..66L}
}

@ARTICLE{2018ApJ...863...67G,
       author = {{Gossage}, Seth and {Conroy}, Charlie and {Dotter}, Aaron and {Choi}, Jieun and {Rosenfield}, Philip and {Cargile}, Philip and {Dolphin}, Andrew},
        title = "{Age Determinations of the Hyades, Praesepe, and Pleiades via MESA Models with Rotation}",
      journal = {\apj},
         year = 2018,
        month = aug,
       volume = {863},
       number = {1},
          eid = {67},
        pages = {67},
          doi = {10.3847/1538-4357/aad0a0},
archivePrefix = {arXiv},
       eprint = {1804.06441},
 primaryClass = {astro-ph.SR},
       adsurl = {https://ui.adsabs.harvard.edu/abs/2018ApJ...863...67G}
}

@INPROCEEDINGS{2011ASPC..448..841D,
       author = {{Delorme}, P. and {Cameron}, A.~C. and {Hebb}, L. and {Rostron}, J. and {Lister}, T.~A. and {Norton}, A.~J. and {Pollacco}, D. and {West}, R.~G.},
        title = "{Stellar Rotation in the Hyades and Praesepe: Gyrochronology and Braking Timescale}",
    booktitle = {16th Cambridge Workshop on Cool Stars, Stellar Systems, and the Sun},
         year = 2011,
       editor = {{Johns-Krull}, Christopher and {Browning}, Matthew K. and {West}, Andrew A.},
       series = {Astronomical Society of the Pacific Conference Series},
       volume = {448},
        month = dec,
        pages = {841},
       adsurl = {https://ui.adsabs.harvard.edu/abs/2011ASPC..448..841D}
}

@ARTICLE{2025ApJ...979L...1L,
       author = {{Leiner}, Emily M. and {Gosnell}, Natalie M. and {Geller}, Aaron M. and {Sun}, Meng and {Mathieu}, Robert D. and {Sills}, Alison},
        title = "{The Blue Lurker WOCS 14020: A Long-period Post-common-envelope Binary in M67 Originating from a Merger in a Triple System}",
      journal = {\apjl},
         year = 2025,
        month = jan,
       volume = {979},
       number = {1},
          eid = {L1},
        pages = {L1},
          doi = {10.3847/2041-8213/ad9d0c},
       adsurl = {https://ui.adsabs.harvard.edu/abs/2025ApJ...979L...1L}
}

@ARTICLE{2022MNRAS.517.2867H,
       author = {{Hernandez}, M.~S. and {Schreiber}, M.~R. and {Parsons}, S.~G. and {G{\"a}nsicke}, B.~T. and {Toloza}, O. and {Zorotovic}, M. and {Raddi}, R. and {Rebassa-Mansergas}, A. and {Ren}, J.~J.},
        title = "{The white dwarf binary pathways survey - VIII. A post-common envelope binary with a massive white dwarf and an active G-type secondary star}",
      journal = {\mnras},
         year = 2022,
        month = dec,
       volume = {517},
       number = {2},
        pages = {2867-2875},
          doi = {10.1093/mnras/stac2837},
archivePrefix = {arXiv},
       eprint = {2209.15591},
 primaryClass = {astro-ph.SR},
       adsurl = {https://ui.adsabs.harvard.edu/abs/2022MNRAS.517.2867H}
}

@ARTICLE{2022MNRAS.512.1843H,
       author = {{Hernandez}, M.~S. and {Schreiber}, M.~R. and {Parsons}, S.~G. and {G{\"a}nsicke}, B.~T. and {Toloza}, O. and {Tovmassian}, G. and {Zorotovic}, M. and {Lagos}, F. and {Raddi}, R. and {Rebassa-Mansergas}, A. and {Ren}, J.~J. and {Tappert}, C.},
        title = "{The white dwarf binary pathways survey - VI. Two close post-common envelope binaries with TESS light curves}",
      journal = {\mnras},
         year = 2022,
        month = may,
       volume = {512},
       number = {2},
        pages = {1843-1856},
          doi = {10.1093/mnras/stac604},
archivePrefix = {arXiv},
       eprint = {2203.01745},
 primaryClass = {astro-ph.SR},
       adsurl = {https://ui.adsabs.harvard.edu/abs/2022MNRAS.512.1843H}
}

@ARTICLE{2021MNRAS.501.1677H,
       author = {{Hernandez}, M.~S. and {Schreiber}, M.~R. and {Parsons}, S.~G. and {G{\"a}nsicke}, B.~T. and {Lagos}, F. and {Raddi}, R. and {Toloza}, O. and {Tovmassian}, G. and {Zorotovic}, M. and {Irawati}, P. and {Past{\'e}n}, E. and {Rebassa-Mansergas}, A. and {Ren}, J.~J. and {Rittipruk}, P. and {Tappert}, C.},
        title = "{The White Dwarf Binary Pathways Survey - IV. Three close white dwarf binaries with G-type secondary stars}",
      journal = {\mnras},
         year = 2021,
        month = feb,
       volume = {501},
       number = {2},
        pages = {1677-1689},
          doi = {10.1093/mnras/staa3815},
archivePrefix = {arXiv},
       eprint = {2012.04683},
 primaryClass = {astro-ph.SR},
       adsurl = {https://ui.adsabs.harvard.edu/abs/2021MNRAS.501.1677H}
}

@ARTICLE{2011A&A...536L...3Z,
       author = {{Zorotovic}, M. and {Schreiber}, M.~R. and {G{\"a}nsicke}, B.~T. and {Rebassa-Mansergas}, A. and {Nebot G{\'o}mez-Mor{\'a}n}, A. and {Southworth}, J. and {Schwope}, A.~D. and {Pyrzas}, S. and {Rodr{\'\i}guez-Gil}, P. and {Schmidtobreick}, L. and {Schwarz}, R. and {Tappert}, C. and {Toloza}, O. and {Vogt}, N.},
        title = "{Post common envelope binaries from SDSS. XIII. Mass dependencies of the orbital period distribution}",
      journal = {\aap},
         year = 2011,
        month = dec,
       volume = {536},
          eid = {L3},
        pages = {L3},
          doi = {10.1051/0004-6361/201117803},
archivePrefix = {arXiv},
       eprint = {1111.2512},
 primaryClass = {astro-ph.SR},
       adsurl = {https://ui.adsabs.harvard.edu/abs/2011A&A...536L...3Z}
}

@ARTICLE{2010A&A...513L...7S,
       author = {{Schreiber}, M.~R. and {G{\"a}nsicke}, B.~T. and {Rebassa-Mansergas}, A. and {Nebot Gomez-Moran}, A. and {Southworth}, J. and {Schwope}, A.~D. and {M{\"u}ller}, M. and {Papadaki}, C. and {Pyrzas}, S. and {Rabitz}, A. and {Rodr{\'\i}guez-Gil}, P. and {Schmidtobreick}, L. and {Schwarz}, R. and {Tappert}, C. and {Toloza}, O. and {Vogel}, J. and {Zorotovic}, M.},
        title = "{Post common envelope binaries from SDSS. VIII. Evidence for disrupted magnetic braking}",
      journal = {\aap},
         year = 2010,
        month = apr,
       volume = {513},
          eid = {L7},
        pages = {L7},
          doi = {10.1051/0004-6361/201013990},
       adsurl = {https://ui.adsabs.harvard.edu/abs/2010A&A...513L...7S}
}

@INPROCEEDINGS{2009JPhCS.172a2024S,
       author = {{Schreiber}, M.~R. and {Gaensicke}, B.~T. and {Zorotovic}, M. and {Rebassa-Mansergas}, A. and {Nebot Gomez-Moran}, A. and {Southworth}, J. and {Schwope}, A.~D. and {Pyrzas}, S. and {Tappert}, C. and {Schmidtobreick}, L.},
        title = "{White dwarf post common envelope binaries from the SDSS}",
    booktitle = {Journal of Physics Conference Series},
         year = 2009,
       series = {Journal of Physics Conference Series},
       volume = {172},
        month = jun,
    publisher = {IOP},
          eid = {012024},
        pages = {012024},
          doi = {10.1088/1742-6596/172/1/012024},
       adsurl = {https://ui.adsabs.harvard.edu/abs/2009JPhCS.172a2024S}
}

@ARTICLE{2018MNRAS.476.2556Q,
       author = {{Queiroz}, A.~B.~A. and {Anders}, F. and {Santiago}, B.~X. and {Chiappini}, C. and {Steinmetz}, M. and {Dal Ponte}, M. and {Stassun}, K.~G. and {da Costa}, L.~N. and {Maia}, M.~A.~G. and {Crestani}, J. and {Beers}, T.~C. and {Fern{\'a}ndez-Trincado}, J.~G. and {Garc{\'\i}a-Hern{\'a}ndez}, D.~A. and {Roman-Lopes}, A. and {Zamora}, O.},
        title = "{StarHorse: a Bayesian tool for determining stellar masses, ages, distances, and extinctions for field stars}",
      journal = {\mnras},
         year = 2018,
        month = may,
       volume = {476},
       number = {2},
        pages = {2556-2583},
          doi = {10.1093/mnras/sty330},
archivePrefix = {arXiv},
       eprint = {1710.09970},
 primaryClass = {astro-ph.IM},
       adsurl = {https://ui.adsabs.harvard.edu/abs/2018MNRAS.476.2556Q}
}

@ARTICLE{2020A&A...638A..76Q,
       author = {{Queiroz}, A.~B.~A. and {Anders}, F. and {Chiappini}, C. and {Khalatyan}, A. and {Santiago}, B.~X. and {Steinmetz}, M. and {Valentini}, M. and {Miglio}, A. and {Bossini}, D. and {Barbuy}, B. and {Minchev}, I. and {Minniti}, D. and {Garc{\'\i}a Hern{\'a}ndez}, D.~A. and {Schultheis}, M. and {Beaton}, R.~L. and {Beers}, T.~C. and {Bizyaev}, D. and {Brownstein}, J.~R. and {Cunha}, K. and {Fern{\'a}ndez-Trincado}, J.~G. and {Frinchaboy}, P.~M. and {Lane}, R.~R. and {Majewski}, S.~R. and {Nataf}, D. and {Nitschelm}, C. and {Pan}, K. and {Roman-Lopes}, A. and {Sobeck}, J.~S. and {Stringfellow}, G. and {Zamora}, O.},
        title = "{From the bulge to the outer disc: StarHorse stellar parameters, distances, and extinctions for stars in APOGEE DR16 and other spectroscopic surveys}",
      journal = {\aap},
         year = 2020,
        month = jun,
       volume = {638},
          eid = {A76},
        pages = {A76},
          doi = {10.1051/0004-6361/201937364},
archivePrefix = {arXiv},
       eprint = {1912.09778},
 primaryClass = {astro-ph.GA},
       adsurl = {https://ui.adsabs.harvard.edu/abs/2020A&A...638A..76Q}
}

@ARTICLE{2010MmSAI..81..921K,
       author = {{Koester}, D.},
        title = "{White dwarf spectra and atmosphere models}",
      journal = {\memsai},
         year = 2010,
        month = jan,
       volume = {81},
        pages = {921-931},
       adsurl = {https://ui.adsabs.harvard.edu/abs/2010MmSAI..81..921K}
}

@misc{2023yCat.3286....0A,
       author = {{Abdurro'Uf} and {et al.}},
        title = "{VizieR Online Data Catalog: APOGEE-2 DR17 final allStar catalog (Abdurro'uf+, 2022)}",
 howpublished = {VizieR On-line Data Catalog: III/286.  Originally published in: 2022ApJS..259...35A},
         year = 2023,
        month = nov,
          eid = {III/286},
       adsurl = {https://ui.adsabs.harvard.edu/abs/2023yCat.3286....0A}
}

@ARTICLE{1967ZA.....65...89L,
       author = {{Lucy}, L.~B.},
        title = "{Gravity-Darkening for Stars with Convective Envelopes}",
      journal = {\zap},
         year = 1967,
        month = jan,
       volume = {65},
        pages = {89},
       adsurl = {https://ui.adsabs.harvard.edu/abs/1967ZA.....65...89L}
}

@ARTICLE{1924MNRAS..84..702V,
       author = {{von Zeipel}, H.},
        title = "{Radiative equilibrium of a double-star system with nearly spherical components}",
      journal = {\mnras},
         year = 1924,
        month = jun,
       volume = {84},
        pages = {702},
          doi = {10.1093/mnras/84.9.702},
       adsurl = {https://ui.adsabs.harvard.edu/abs/1924MNRAS..84..702V}
}

@ARTICLE{1976Ap&SS..39..447L,
       author = {{Lomb}, N.~R.},
        title = "{Least-Squares Frequency Analysis of Unequally Spaced Data}",
      journal = {\apss},
         year = 1976,
        month = feb,
       volume = {39},
       number = {2},
        pages = {447-462},
          doi = {10.1007/BF00648343},
       adsurl = {https://ui.adsabs.harvard.edu/abs/1976Ap&SS..39..447L}
}

@ARTICLE{1982ApJ...263..835S,
       author = {{Scargle}, J.~D.},
        title = "{Studies in astronomical time series analysis. II. Statistical aspects of spectral analysis of unevenly spaced data.}",
      journal = {\apj},
         year = 1982,
        month = dec,
       volume = {263},
        pages = {835-853},
          doi = {10.1086/160554},
       adsurl = {https://ui.adsabs.harvard.edu/abs/1982ApJ...263..835S}
}

@ARTICLE{2020AJ....160..269M,
       author = {{Melbourne}, Katherine and {Youngblood}, Allison and {France}, Kevin and {Froning}, C.~S. and {Pineda}, J. Sebastian and {Shkolnik}, Evgenya L. and {Wilson}, David J. and {Wood}, Brian E. and {Basu}, Sarbani and {Roberge}, Aki and {Schlieder}, Joshua E. and {Cauley}, P. Wilson and {Loyd}, R.~O. Parke and {Newton}, Elisabeth R. and {Schneider}, Adam and {Arulanantham}, Nicole and {Berta-Thompson}, Zachory and {Brown}, Alexander and {Buccino}, Andrea P. and {Kempton}, Eliza and {Linsky}, Jeffrey L. and {Logsdon}, Sarah E. and {Mauas}, Pablo and {Pagano}, Isabella and {Peacock}, Sarah and {Redfield}, Seth and {Rugheimer}, Sarah and {Schneider}, P. Christian and {Teal}, D.~J. and {Tian}, Feng and {Tilipman}, Dennis and {Vieytes}, Mariela},
        title = "{Estimating the Ultraviolet Emission of M Dwarfs with Exoplanets from Ca II and H{\ensuremath{\alpha}}}",
      journal = {\aj},
         year = 2020,
        month = dec,
       volume = {160},
       number = {6},
          eid = {269},
        pages = {269},
          doi = {10.3847/1538-3881/abbf5c},
archivePrefix = {arXiv},
       eprint = {2009.07869},
 primaryClass = {astro-ph.SR},
       adsurl = {https://ui.adsabs.harvard.edu/abs/2020AJ....160..269M}
}

@ARTICLE{2020A&A...641A.136W,
       author = {{Webb}, N.~A. and {Coriat}, M. and {Traulsen}, I. and {Ballet}, J. and {Motch}, C. and {Carrera}, F.~J. and {Koliopanos}, F. and {Authier}, J. and {de la Calle}, I. and {Ceballos}, M.~T. and {Colomo}, E. and {Chuard}, D. and {Freyberg}, M. and {Garcia}, T. and {Kolehmainen}, M. and {Lamer}, G. and {Lin}, D. and {Maggi}, P. and {Michel}, L. and {Page}, C.~G. and {Page}, M.~J. and {Perea-Calderon}, J.~V. and {Pineau}, F. -X. and {Rodriguez}, P. and {Rosen}, S.~R. and {Santos Lleo}, M. and {Saxton}, R.~D. and {Schwope}, A. and {Tom{\'a}s}, L. and {Watson}, M.~G. and {Zakardjian}, A.},
        title = "{The XMM-Newton serendipitous survey. IX. The fourth XMM-Newton serendipitous source catalogue}",
      journal = {\aap},
         year = 2020,
        month = sep,
       volume = {641},
          eid = {A136},
        pages = {A136},
          doi = {10.1051/0004-6361/201937353},
archivePrefix = {arXiv},
       eprint = {2007.02899},
 primaryClass = {astro-ph.HE},
       adsurl = {https://ui.adsabs.harvard.edu/abs/2020A&A...641A.136W}
}

@ARTICLE{2024MNRAS.529.4840G,
       author = {{Garbutt}, J.~A. and {Parsons}, S.~G. and {Toloza}, O. and {G{\"a}nsicke}, B.~T. and {Hernandez}, M.~S. and {Koester}, D. and {Lagos}, F. and {Raddi}, R. and {Rebassa-Mansergas}, A. and {Ren}, J.~J. and {Schreiber}, M.~R. and {Zorotovic}, M.},
        title = "{The white dwarf binary pathways survey - X. Gaia orbits for known UV excess binaries}",
      journal = {\mnras},
         year = 2024,
        month = apr,
       volume = {529},
       number = {4},
        pages = {4840-4855},
          doi = {10.1093/mnras/stae807},
archivePrefix = {arXiv},
       eprint = {2403.07985},
 primaryClass = {astro-ph.SR},
       adsurl = {https://ui.adsabs.harvard.edu/abs/2024MNRAS.529.4840G}
}

@misc{2020yCat.1350....0G,
       author = {{Gaia Collaboration}},
        title = "{VizieR Online Data Catalog: Gaia EDR3 (Gaia Collaboration, 2020)}",
 howpublished = {VizieR On-line Data Catalog: I/350.  Originally published in: 2021A\&A...649A...1G},
         year = 2020,
        month = nov,
          eid = {I/350},
          doi = {10.26093/cds/vizier.1350},
       adsurl = {https://ui.adsabs.harvard.edu/abs/2020yCat.1350....0G}
}

@ARTICLE{2017MNRAS.472.4193R,
       author = {{Rebassa-Mansergas}, A. and {Ren}, J.~J. and {Irawati}, P. and {Garc{\'\i}a-Berro}, E. and {Parsons}, S.~G. and {Schreiber}, M.~R. and {G{\"a}nsicke}, B.~T. and {Rodr{\'\i}guez-Gil}, P. and {Liu}, X. and {Manser}, C. and {Nevado}, S.~P. and {Jim{\'e}nez-Ibarra}, F. and {Costero}, R. and {Echevarr{\'\i}a}, J. and {Michel}, R. and {Zorotovic}, M. and {Hollands}, M. and {Han}, Z. and {Luo}, A. and {Villaver}, E. and {Kong}, X.},
        title = "{The white dwarf binary pathways survey - II. Radial velocities of 1453 FGK stars with white dwarf companions from LAMOST DR 4}",
      journal = {\mnras},
         year = 2017,
        month = dec,
       volume = {472},
       number = {4},
        pages = {4193-4203},
          doi = {10.1093/mnras/stx2259},
archivePrefix = {arXiv},
       eprint = {1708.09480},
 primaryClass = {astro-ph.SR},
       adsurl = {https://ui.adsabs.harvard.edu/abs/2017MNRAS.472.4193R}
}

@ARTICLE{2023ApJ...954..134K,
       author = {{Kerr}, Ronan and {Kraus}, Adam L. and {Rizzuto}, Aaron C.},
        title = "{SPYGLASS. IV. New Stellar Survey of Recent Star Formation within 1 kpc}",
      journal = {\apj},
         year = 2023,
        month = sep,
       volume = {954},
       number = {2},
          eid = {134},
        pages = {134},
          doi = {10.3847/1538-4357/ace5b3},
archivePrefix = {arXiv},
       eprint = {2306.08150},
 primaryClass = {astro-ph.GA},
       adsurl = {https://ui.adsabs.harvard.edu/abs/2023ApJ...954..134K}
}

@ARTICLE{Breivik2020,
       author = {{Breivik}, Katelyn and {Coughlin}, Scott and {Zevin}, Michael and {Rodriguez}, Carl L. and {Kremer}, Kyle and {Ye}, Claire S. and {Andrews}, Jeff J. and {Kurkowski}, Michael and {Digman}, Matthew C. and {Larson}, Shane L. and {Rasio}, Frederic A.},
        title = "{COSMIC Variance in Binary Population Synthesis}",
      journal = {\apj},
         year = 2020,
        month = jul,
       volume = {898},
       number = {1},
          eid = {71},
        pages = {71},
          doi = {10.3847/1538-4357/ab9d85},
archivePrefix = {arXiv},
       eprint = {1911.00903},
 primaryClass = {astro-ph.HE},
       adsurl = {https://ui.adsabs.harvard.edu/abs/2020ApJ...898...71B}
}

@ARTICLE{2019MNRAS.490.3740N,
       author = {{Neijssel}, Coenraad J. and {Vigna-G{\'o}mez}, Alejandro and {Stevenson}, Simon and {Barrett}, Jim W. and {Gaebel}, Sebastian M. and {Broekgaarden}, Floor S. and {de Mink}, Selma E. and {Sz{\'e}csi}, Dorottya and {Vinciguerra}, Serena and {Mandel}, Ilya},
        title = "{The effect of the metallicity-specific star formation history on double compact object mergers}",
      journal = {\mnras},
         year = 2019,
        month = dec,
       volume = {490},
       number = {3},
        pages = {3740-3759},
          doi = {10.1093/mnras/stz2840},
archivePrefix = {arXiv},
       eprint = {1906.08136},
 primaryClass = {astro-ph.SR},
       adsurl = {https://ui.adsabs.harvard.edu/abs/2019MNRAS.490.3740N}
}

@ARTICLE{2024PASP..136h4202Y,
       author = {{Yamaguchi}, Natsuko and {El-Badry}, Kareem and {Rees}, Natalie R. and {Shahaf}, Sahar and {Mazeh}, Tsevi and {Andrae}, Ren{\'r}},
        title = "{Wide Post-common Envelope Binaries from Gaia: Orbit Validation and Formation Models}",
      journal = {\pasp},
         year = 2024,
        month = aug,
       volume = {136},
       number = {8},
          eid = {084202},
        pages = {084202},
          doi = {10.1088/1538-3873/ad6809},
archivePrefix = {arXiv},
       eprint = {2405.06020},
 primaryClass = {astro-ph.SR},
       adsurl = {https://ui.adsabs.harvard.edu/abs/2024PASP..136h4202Y}
}

@ARTICLE{2005A&A...441..689A,
       author = {{Althaus}, L.~G. and {Garc{\'\i}a-Berro}, E. and {Isern}, J. and {C{\'o}rsico}, A.~H.},
        title = "{Mass-radius relations for massive white dwarf stars}",
      journal = {\aap},
         year = 2005,
        month = oct,
       volume = {441},
       number = {2},
        pages = {689-694},
          doi = {10.1051/0004-6361:20052996},
archivePrefix = {arXiv},
       eprint = {astro-ph/0507559},
 primaryClass = {astro-ph},
       adsurl = {https://ui.adsabs.harvard.edu/abs/2005A&A...441..689A}
}

@ARTICLE{2022MNRAS.517.3181G,
       author = {{Gonz{\'a}lez-Bol{\'\i}var}, Miguel and {De Marco}, Orsola and {Lau}, Mike Y.~M. and {Hirai}, Ryosuke and {Price}, Daniel J.},
        title = "{Common envelope binary interaction simulations between a thermally pulsating AGB star and a low mass companion}",
      journal = {\mnras},
         year = 2022,
        month = dec,
       volume = {517},
       number = {3},
        pages = {3181-3199},
          doi = {10.1093/mnras/stac2301},
archivePrefix = {arXiv},
       eprint = {2205.09749},
 primaryClass = {astro-ph.SR},
       adsurl = {https://ui.adsabs.harvard.edu/abs/2022MNRAS.517.3181G}
}

@ARTICLE{2016ApJ...823..102C,
       author = {{Choi}, Jieun and {Dotter}, Aaron and {Conroy}, Charlie and {Cantiello}, Matteo and {Paxton}, Bill and {Johnson}, Benjamin D.},
        title = "{Mesa Isochrones and Stellar Tracks (MIST). I. Solar-scaled Models}",
      journal = {\apj},
         year = 2016,
        month = jun,
       volume = {823},
       number = {2},
          eid = {102},
        pages = {102},
          doi = {10.3847/0004-637X/823/2/102},
archivePrefix = {arXiv},
       eprint = {1604.08592},
 primaryClass = {astro-ph.SR},
       adsurl = {https://ui.adsabs.harvard.edu/abs/2016ApJ...823..102C}
}

@ARTICLE{2016ApJS..222....8D,
       author = {{Dotter}, Aaron},
        title = "{MESA Isochrones and Stellar Tracks (MIST) 0: Methods for the Construction of Stellar Isochrones}",
      journal = {\apjs},
         year = 2016,
        month = jan,
       volume = {222},
       number = {1},
          eid = {8},
        pages = {8},
          doi = {10.3847/0067-0049/222/1/8},
archivePrefix = {arXiv},
       eprint = {1601.05144},
 primaryClass = {astro-ph.SR},
       adsurl = {https://ui.adsabs.harvard.edu/abs/2016ApJS..222....8D}
}

@ARTICLE{2012MNRAS.427..127B,
       author = {{Bressan}, Alessandro and {Marigo}, Paola and {Girardi}, L{\'e}o. and {Salasnich}, Bernardo and {Dal Cero}, Claudia and {Rubele}, Stefano and {Nanni}, Ambra},
        title = "{PARSEC: stellar tracks and isochrones with the PAdova and TRieste Stellar Evolution Code}",
      journal = {\mnras},
         year = 2012,
        month = nov,
       volume = {427},
       number = {1},
        pages = {127-145},
          doi = {10.1111/j.1365-2966.2012.21948.x},
archivePrefix = {arXiv},
       eprint = {1208.4498},
 primaryClass = {astro-ph.SR},
       adsurl = {https://ui.adsabs.harvard.edu/abs/2012MNRAS.427..127B}
}

@ARTICLE{2025arXiv250406252H,
       author = {{Herrera-Urquieta}, A. and {Leigh}, N. and {Pinto}, J. and {D{\'\i}az-Cerda}, G. and {Grondin}, S.~M. and {Webb}, J.~J. and {Mathieu}, R. and {Ryu}, T. and {Geller}, A. and {Kounkel}, M. and {Toonen}, S. and {Vilaxa-Campos}, M.},
        title = "{A systematic method to identify runaways from star clusters produced from single-binary interactions: A case study of M67}",
      journal = {arXiv e-prints},
         year = 2025,
        month = apr,
          eid = {arXiv:2504.06252},
        pages = {arXiv:2504.06252},
          doi = {10.48550/arXiv.2504.06252},
archivePrefix = {arXiv},
       eprint = {2504.06252},
 primaryClass = {astro-ph.SR},
       adsurl = {https://ui.adsabs.harvard.edu/abs/2025arXiv250406252H}
}

@ARTICLE{Horch2009,
       author = {{Horch}, Elliott P. and {Veillette}, Daniel R. and {Baena Gall{\'e}}, Roberto and {Shah}, Sagar C. and {O'Rielly}, Grant V. and {van Altena}, William F.},
        title = "{Observations of Binary Stars with the Differential Speckle Survey Instrument. I. Instrument Description and First Results}",
      journal = {\aj},
         year = 2009,
        month = jun,
       volume = {137},
       number = {6},
        pages = {5057-5067},
          doi = {10.1088/0004-6256/137/6/5057},
       adsurl = {https://ui.adsabs.harvard.edu/abs/2009AJ....137.5057H}
}

@ARTICLE{Horch2011a,
       author = {{Horch}, Elliott P. and {Gomez}, Shamilia C. and {Sherry}, William H. and {Howell}, Steve B. and {Ciardi}, David R. and {Anderson}, Lisa M. and {van Altena}, William F.},
        title = "{Observations of Binary Stars with the Differential Speckle Survey Instrument. II. Hipparcos Stars Observed in 2010 January and June}",
      journal = {\aj},
         year = 2011,
        month = feb,
       volume = {141},
       number = {2},
          eid = {45},
        pages = {45},
          doi = {10.1088/0004-6256/141/2/45},
       adsurl = {https://ui.adsabs.harvard.edu/abs/2011AJ....141...45H}
}

@ARTICLE{Horch2021,
       author = {{Horch}, Elliott P. and {Broderick}, Kyle G. and {Casetti-Dinescu}, Dana I. and {Henry}, Todd J. and {Fekel}, Francis C. and {Muterspaugh}, Matthew W. and {Willmarth}, Daryl W. and {Winters}, Jennifer G. and {van Belle}, Gerard T. and {Clark}, Catherine A. and {Everett}, Mark E.},
        title = "{Observations with the Differential Speckle Survey Instrument. X. Preliminary Orbits of K-dwarf Binaries and Other Stars}",
      journal = {\aj},
         year = 2021,
        month = jun,
       volume = {161},
       number = {6},
          eid = {295},
        pages = {295},
          doi = {10.3847/1538-3881/abf9a8},
archivePrefix = {arXiv},
       eprint = {2104.07760},
 primaryClass = {astro-ph.SR},
       adsurl = {https://ui.adsabs.harvard.edu/abs/2021AJ....161..295H}
}

@ARTICLE{Davidson2024,
       author = {{Davidson}, James W. and {Horch}, Elliott P. and {Majewski}, Steven R. and {Fagan}, Evan and {Shea}, Melissa A. and {Sutherland}, Torrie and {Wilson}, Robert F. and {Lesley}, D. Xavier and {Pellegrino}, Richard A. and {Leonard}, Jonathan P. and {Wilson}, John C. and {Chanover}, Nancy J. and {Dow}, Peter and {Henry}, Todd J. and {Ketzeback}, William and {McDonald}, Devin and {McMillan}, Russet and {Dembicky}, Jack and {DeColibus}, Riley A. and {Gray}, Candace and {Townsend}, Amanda},
        title = "{Observations with the Differential Speckle Survey Instrument. XI. First Year of Observations from Apache Point Observatory}",
      journal = {\aj},
         year = 2024,
        month = mar,
       volume = {167},
       number = {3},
          eid = {117},
        pages = {117},
          doi = {10.3847/1538-3881/ad1ff6},
       adsurl = {https://ui.adsabs.harvard.edu/abs/2024AJ....167..117D}
}

@misc{2020yCat..36410157C,
       author = {{Claret}, A. and {Cukanovaite}, E. and {Burdge}, K. and {Tremblay}, P. -E. and {Parsons}, S. and {Marsh}, T.~R.},
        title = "{VizieR Online Data Catalog: Limb-darkening coefficients for white dwarfs (Claret+, 2020)}",
 howpublished = {VizieR On-line Data Catalog: J/A+A/641/A157. Originally published in: 2020A\&A...641A.157C},
         year = 2020,
        month = jul,
          eid = {J/A+A/641/A157},
          doi = {10.26093/cds/vizier.36410157},
       adsurl = {https://ui.adsabs.harvard.edu/abs/2020yCat..36410157C}
}

@ARTICLE{2008ApJ...685..553S,
       author = {{Shen}, Yue and {Turner}, Edwin L.},
        title = "{On the Eccentricity Distribution of Exoplanets from Radial Velocity Surveys}",
      journal = {\apj},
         year = 2008,
        month = sep,
       volume = {685},
       number = {1},
        pages = {553-559},
          doi = {10.1086/590548},
archivePrefix = {arXiv},
       eprint = {0806.0032},
 primaryClass = {astro-ph},
       adsurl = {https://ui.adsabs.harvard.edu/abs/2008ApJ...685..553S}
}

@ARTICLE{1993MNRAS.262..277W,
       author = {{Wonnacott}, D. and {Kellett}, B.~J. and {Stickland}, D.~J.},
        title = "{IK Peg : a nearby, short-period, Sirius-like system.}",
      journal = {\mnras},
         year = 1993,
        month = may,
       volume = {262},
        pages = {277-284},
          doi = {10.1093/mnras/262.2.277},
       adsurl = {https://ui.adsabs.harvard.edu/abs/1993MNRAS.262..277W}
}

@ARTICLE{2010A&A...520A..86Z,
       author = {{Zorotovic}, M. and {Schreiber}, M.~R. and {G{\"a}nsicke}, B.~T. and {Nebot G{\'o}mez-Mor{\'a}n}, A.},
        title = "{Post-common-envelope binaries from SDSS. IX: Constraining the common-envelope efficiency}",
      journal = {\aap},
         year = 2010,
        month = sep,
       volume = {520},
          eid = {A86},
        pages = {A86},
          doi = {10.1051/0004-6361/200913658},
archivePrefix = {arXiv},
       eprint = {1006.1621},
 primaryClass = {astro-ph.SR},
       adsurl = {https://ui.adsabs.harvard.edu/abs/2010A&A...520A..86Z}
}

@ARTICLE{1987ApJ...318..794H,
       author = {{Hjellming}, Michael S. and {Webbink}, Ronald F.},
        title = "{Thresholds for Rapid Mass Transfer in Binary System. I. Polytropic Models}",
      journal = {\apj},
         year = 1987,
        month = jul,
       volume = {318},
        pages = {794},
          doi = {10.1086/165412},
       adsurl = {https://ui.adsabs.harvard.edu/abs/1987ApJ...318..794H}
}

@INPROCEEDINGS{2010AIPC.1314...53G,
       author = {{Ge}, Hongwei and {Webbink}, Ronald F. and {Chen}, Xuefei and {Han}, Zhanwen},
        title = "{Stellar Adiabatic Mass Loss in Binary Stars}",
    booktitle = {International Conference on Binaries: in celebration of Ron Webbink's 65th Birthday},
         year = 2010,
       editor = {{Kalogera}, Vicky and {van der Sluys}, Marc},
       series = {American Institute of Physics Conference Series},
       volume = {1314},
        month = dec,
    publisher = {AIP},
        pages = {53-54},
          doi = {10.1063/1.3536410},
       adsurl = {https://ui.adsabs.harvard.edu/abs/2010AIPC.1314...53G}
}

@ARTICLE{2015ApJ...812...40G,
       author = {{Ge}, Hongwei and {Webbink}, Ronald F. and {Chen}, Xuefei and {Han}, Zhanwen},
        title = "{Adiabatic Mass Loss in Binary Stars. II. From Zero-age Main Sequence to the Base of the Giant Branch}",
      journal = {\apj},
         year = 2015,
        month = oct,
       volume = {812},
       number = {1},
          eid = {40},
        pages = {40},
          doi = {10.1088/0004-637X/812/1/40},
archivePrefix = {arXiv},
       eprint = {1507.04843},
 primaryClass = {astro-ph.SR},
       adsurl = {https://ui.adsabs.harvard.edu/abs/2015ApJ...812...40G}
}

@ARTICLE{2020ApJ...899..132G,
       author = {{Ge}, Hongwei and {Webbink}, Ronald F. and {Chen}, Xuefei and {Han}, Zhanwen},
        title = "{Adiabatic Mass Loss in Binary Stars. III. From the Base of the Red Giant Branch to the Tip of the Asymptotic Giant Branch}",
      journal = {\apj},
         year = 2020,
        month = aug,
       volume = {899},
       number = {2},
          eid = {132},
        pages = {132},
          doi = {10.3847/1538-4357/aba7b7},
archivePrefix = {arXiv},
       eprint = {2007.09848},
 primaryClass = {astro-ph.SR},
       adsurl = {https://ui.adsabs.harvard.edu/abs/2020ApJ...899..132G}
}

@ARTICLE{2023A&A...669A..45T,
       author = {{Temmink}, K.~D. and {Pols}, O.~R. and {Justham}, S. and {Istrate}, A.~G. and {Toonen}, S.},
        title = "{Coping with loss. Stability of mass transfer from post-main-sequence donor stars}",
      journal = {\aap},
         year = 2023,
        month = jan,
       volume = {669}, 
          eid = {A45},
        pages = {A45},
          doi = {10.1051/0004-6361/202244137},
archivePrefix = {arXiv},
       eprint = {2209.12707},
 primaryClass = {astro-ph.SR},
       adsurl = {https://ui.adsabs.harvard.edu/abs/2023A&A...669A..45T}
}

@ARTICLE{2007MNRAS.382.1377R,
       author = {{Rebassa-Mansergas}, A. and {G{\"a}nsicke}, B.~T. and {Rodr{\'\i}guez-Gil}, P. and {Schreiber}, M.~R. and {Koester}, D.},
        title = "{Post-common-envelope binaries from SDSS - I. 101 white dwarf main-sequence binaries with multiple Sloan Digital Sky Survey spectroscopy}",
      journal = {\mnras},
         year = 2007,
        month = dec,
       volume = {382},
       number = {4},
        pages = {1377-1393},
          doi = {10.1111/j.1365-2966.2007.12288.x},
archivePrefix = {arXiv},
       eprint = {0707.4107},
 primaryClass = {astro-ph},
       adsurl = {https://ui.adsabs.harvard.edu/abs/2007MNRAS.382.1377R}
}

@ARTICLE{1995MNRAS.273..731R,
       author = {{Rappaport}, S. and {Podsiadlowski}, Ph. and {Joss}, P.~C. and {Di Stefano}, R. and {Han}, Z.},
        title = "{The relation between white dwarf mass and orbital period in wide binary radio pulsars}",
      journal = {\mnras},
         year = 1995,
        month = apr,
       volume = {273},
       number = {3},
        pages = {731-741},
          doi = {10.1093/mnras/273.3.731},
       adsurl = {https://ui.adsabs.harvard.edu/abs/1995MNRAS.273..731R}
}

@ARTICLE{2013A&A...558A..33A,
       author = {{Astropy Collaboration} and {Robitaille}, Thomas P. and {Tollerud}, Erik J. and {Greenfield}, Perry and {Droettboom}, Michael and {Bray}, Erik and {Aldcroft}, Tom and {Davis}, Matt and {Ginsburg}, Adam and {Price-Whelan}, Adrian M. and {Kerzendorf}, Wolfgang E. and {Conley}, Alexander and {Crighton}, Neil and {Barbary}, Kyle and {Muna}, Demitri and {Ferguson}, Henry and {Grollier}, Fr{\'e}d{\'e}ric and {Parikh}, Madhura M. and {Nair}, Prasanth H. and {Unther}, Hans M. and {Deil}, Christoph and {Woillez}, Julien and {Conseil}, Simon and {Kramer}, Roban and {Turner}, James E.~H. and {Singer}, Leo and {Fox}, Ryan and {Weaver}, Benjamin A. and {Zabalza}, Victor and {Edwards}, Zachary I. and {Azalee Bostroem}, K. and {Burke}, D.~J. and {Casey}, Andrew R. and {Crawford}, Steven M. and {Dencheva}, Nadia and {Ely}, Justin and {Jenness}, Tim and {Labrie}, Kathleen and {Lim}, Pey Lian and {Pierfederici}, Francesco and {Pontzen}, Andrew and {Ptak}, Andy and {Refsdal}, Brian and {Servillat}, Mathieu and {Streicher}, Ole},
        title = "{Astropy: A community Python package for astronomy}",
      journal = {\aap},
         year = 2013,
        month = oct,
       volume = {558},
          eid = {A33},
        pages = {A33},
          doi = {10.1051/0004-6361/201322068},
archivePrefix = {arXiv},
       eprint = {1307.6212},
 primaryClass = {astro-ph.IM},
       adsurl = {https://ui.adsabs.harvard.edu/abs/2013A&A...558A..33A}
}

@ARTICLE{2018AJ....156..123A,
       author = {{Astropy Collaboration} and {Price-Whelan}, A.~M. and {Sip{\H{o}}cz}, B.~M. and {G{\"u}nther}, H.~M. and {Lim}, P.~L. and {Crawford}, S.~M. and {Conseil}, S. and {Shupe}, D.~L. and {Craig}, M.~W. and {Dencheva}, N. and {Ginsburg}, A. and {VanderPlas}, J.~T. and {Bradley}, L.~D. and {P{\'e}rez-Su{\'a}rez}, D. and {de Val-Borro}, M. and {Aldcroft}, T.~L. and {Cruz}, K.~L. and {Robitaille}, T.~P. and {Tollerud}, E.~J. and {Ardelean}, C. and {Babej}, T. and {Bach}, Y.~P. and {Bachetti}, M. and {Bakanov}, A.~V. and {Bamford}, S.~P. and {Barentsen}, G. and {Barmby}, P. and {Baumbach}, A. and {Berry}, K.~L. and {Biscani}, F. and {Boquien}, M. and {Bostroem}, K.~A. and {Bouma}, L.~G. and {Brammer}, G.~B. and {Bray}, E.~M. and {Breytenbach}, H. and {Buddelmeijer}, H. and {Burke}, D.~J. and {Calderone}, G. and {Cano Rodr{\'\i}guez}, J.~L. and {Cara}, M. and {Cardoso}, J.~V.~M. and {Cheedella}, S. and {Copin}, Y. and {Corrales}, L. and {Crichton}, D. and {D'Avella}, D. and {Deil}, C. and {Depagne}, {\'E}. and {Dietrich}, J.~P. and {Donath}, A. and {Droettboom}, M. and {Earl}, N. and {Erben}, T. and {Fabbro}, S. and {Ferreira}, L.~A. and {Finethy}, T. and {Fox}, R.~T. and {Garrison}, L.~H. and {Gibbons}, S.~L.~J. and {Goldstein}, D.~A. and {Gommers}, R. and {Greco}, J.~P. and {Greenfield}, P. and {Groener}, A.~M. and {Grollier}, F. and {Hagen}, A. and {Hirst}, P. and {Homeier}, D. and {Horton}, A.~J. and {Hosseinzadeh}, G. and {Hu}, L. and {Hunkeler}, J.~S. and {Ivezi{\'c}}, {\v{Z}}. and {Jain}, A. and {Jenness}, T. and {Kanarek}, G. and {Kendrew}, S. and {Kern}, N.~S. and {Kerzendorf}, W.~E. and {Khvalko}, A. and {King}, J. and {Kirkby}, D. and {Kulkarni}, A.~M. and {Kumar}, A. and {Lee}, A. and {Lenz}, D. and {Littlefair}, S.~P. and {Ma}, Z. and {Macleod}, D.~M. and {Mastropietro}, M. and {McCully}, C. and {Montagnac}, S. and {Morris}, B.~M. and {Mueller}, M. and {Mumford}, S.~J. and {Muna}, D. and {Murphy}, N.~A. and {Nelson}, S. and {Nguyen}, G.~H. and {Ninan}, J.~P. and {N{\"o}the}, M. and {Ogaz}, S. and {Oh}, S. and {Parejko}, J.~K. and {Parley}, N. and {Pascual}, S. and {Patil}, R. and {Patil}, A.~A. and {Plunkett}, A.~L. and {Prochaska}, J.~X. and {Rastogi}, T. and {Reddy Janga}, V. and {Sabater}, J. and {Sakurikar}, P. and {Seifert}, M. and {Sherbert}, L.~E. and {Sherwood-Taylor}, H. and {Shih}, A.~Y. and {Sick}, J. and {Silbiger}, M.~T. and {Singanamalla}, S. and {Singer}, L.~P. and {Sladen}, P.~H. and {Sooley}, K.~A. and {Sornarajah}, S. and {Streicher}, O. and {Teuben}, P. and {Thomas}, S.~W. and {Tremblay}, G.~R. and {Turner}, J.~E.~H. and {Terr{\'o}n}, V. and {van Kerkwijk}, M.~H. and {de la Vega}, A. and {Watkins}, L.~L. and {Weaver}, B.~A. and {Whitmore}, J.~B. and {Woillez}, J. and {Zabalza}, V. and {Astropy Contributors}},
        title = "{The Astropy Project: Building an Open-science Project and Status of the v2.0 Core Package}",
      journal = {\aj},
         year = 2018,
        month = sep,
       volume = {156},
       number = {3},
          eid = {123},
        pages = {123},
          doi = {10.3847/1538-3881/aabc4f},
archivePrefix = {arXiv},
       eprint = {1801.02634},
 primaryClass = {astro-ph.IM},
       adsurl = {https://ui.adsabs.harvard.edu/abs/2018AJ....156..123A}
}

@misc{2018ascl.soft12013L,
       author = {{Lightkurve Collaboration} and {Cardoso}, Jos{\'e} Vin{\'\i}cius de Miranda and {Hedges}, Christina and {Gully-Santiago}, Michael and {Saunders}, Nicholas and {Cody}, Ann Marie and {Barclay}, Thomas and {Hall}, Oliver and {Sagear}, Sheila and {Turtelboom}, Emma and {Zhang}, Johnny and {Tzanidakis}, Andy and {Mighell}, Ken and {Coughlin}, Jeff and {Bell}, Keaton and {Berta-Thompson}, Zach and {Williams}, Peter and {Dotson}, Jessie and {Barentsen}, Geert},
        title = "{Lightkurve: Kepler and TESS time series analysis in Python}",
 howpublished = {Astrophysics Source Code Library, record ascl:1812.013},
         year = 2018,
        month = dec,
          eid = {ascl:1812.013},
       adsurl = {https://ui.adsabs.harvard.edu/abs/2018ascl.soft12013L}
}
\bibliographystyle{aastex701}

\end{document}